\documentclass{aa}  

\usepackage{graphicx}
\usepackage{txfonts}
\begin{document} 

\title{eROSITA all-sky survey - stars and optical loading}

\author{J. Robrade\inst{1}
\and K. Dennerl\inst{2}
\and M.J. Freyberg\inst{2}
\and J.H.M.M. Schmitt\inst{1}
}

\institute{Hamburger Sternwarte, University of Hamburg, Gojenbergsweg 112, 21029 Hamburg, Germany\\
\email{jan.robrade@uni-hamburg.de}
\and Max-Planck-Institut für Extraterrestrische Physik, Gießenbachstraße, 85748 Garching, Germany\\
}
\date{Received..; accepted...}

\abstract
{eROSITA (extended ROentgen Survey with an Imaging Telescope Array) on board the Spectrum-Roentgen-Gamma (SRG) spacecraft has performed the eROSITA All-Sky Survey (eRASS) at X-ray energies.}
{We study the brighter stars to assess the impact of optical loading in the eROSITA all-sky survey, here specifically the eRASS:3. Further we use thermal plasma models to investigate the general properties of stellar eRASS sources and compare optical loading affected stars to coronal sources.}
{We compare data from Gaia DR3, Tycho-2, 2MASS and the SIMBAD database with the eRASS count rates of brighter stars. We analyze X-ray images, light curves and spectra for sources suspected to be affected by optical loading and determine their properties. eROSITA source properties are checked in comparison to genuine coronal sources with generated sets of APEC model spectra, where we determine rate conversion factors and hardness ratios for stellar X-ray emitter.}
{Regarding optical loading we find a clear correlation between optical brightness and observed minimum count rate in the range of about 2\,--\,5 mag (V/G band), which then saturates for very bright sources. While magnitude and color are the main determinants, exposure depth and specific scanning paths also matter. For intrinsic X-ray emitters, optical contamination is likewise observed. We characterize the fake X-ray sources, address pseudo variability and spectral properties. To interpret stellar eRASS data, hardness ratios and count rate conversions for thermal plasma models are provided.}
{The eROSITA data is prone to optical loading, which was studied in greater detail. The optical loading flag for the DR2 is set, if sources are brighter than 5.0~mag (B, V, G) or 3.5~mag (J). It is somewhat dependent on exposure depth, but we expect only minor changes for future data releases.
}


\maketitle


\section{Introduction}

eROSITA \citep{erosita} is the soft X-ray instrument on board the SRG spacecraft \citep{srg}, which performed an all-sky survey in the 0.2\,--\,10.0~keV energy range, the eROSITA all-sky survey (eRASS). In this paper we use primary observations taken from December 2019 until June 2021, i.e. during the first three 0.5~yr all-sky surveys. The data from the surveys are processed individually and in a cumulative fashion. All analyzed data is from the eROSITA\_DE region of the sky, which is basically the galactic western hemisphere. The corresponding eRASS:3 source catalogs (eRASS:3 = eRASS1\,+\,eRASS2\,+\,eRASS3) are published in the context of the data release eROSITA DR2 \citep{DR2}.

The eROSITA instrument uses CCD cameras for the detection of X-ray photons and these are affected by optical loading, i.e. the accumulation of low energy photons and thus charge during the integration time. Even with special filters there is always a trade-off between effective area at X-ray energies and the attenuation of optical light and thus optical blocking is never perfect.
Here we refer to optical loading or contamination, but of course also IR and UV photons can contribute. This is different to the ROSAT all-sky survey \citep[RASS,][]{1RXS, 2RXS}, because ROSAT used a proportional counter detector (PSPC) to collect the survey photons. Overall, while the RASS is basically unaffected by optical loading effects, the eRASS is definitely not.

As counting events caused by optical loading appear identical to 'real' X-rays, such events or sources caused by optical loading are not removed by the standard data analysis systems. Thus in eRASS sources that are near or identified with stars brighter than about 5~mag, optical loading effects are quite likely present. Among the consequences of optical loading are the presence of fake X-ray sources, pseudo X-ray variability and the distortion of real X-ray sources, whereas here the relevance of these effects is governed by the contrast between optical loading signal and intrinsic X-ray brightness.

The goal of this paper is to present presents an in-depth study of the optical loading effects in the eRASS data and to provide a guideline to the general observer for an interpretation of their results. We simulate X-ray data with thermal plasma emission models, determine count rate conversations and hardness ratios and compare them to expectations from optical loading and the observed values. The derived performance parameters and conversions can likewise be used to analyze the eRASS catalog sources and to address the properties of stellar populations.

Our paper is thus structured as follows. In Sect.~\ref{optload} we provide a more detailed overview over optical loading effects in the eRASS data, in Sect.~\ref{obs} we describe the eROSITA X-ray detectors, the employed data and used analysis methods, in Sect.~\ref{res_opt} we present the results of our optical loading studies, in Sect.~\ref{res_cal} we study several eRASS performance parameter and in Sect.~\ref{sum} we discuss and summarize our findings.

\section{eROSITA: Overview of optical loading}
\label{optload}

To address optical loading effects in the eROSITA survey data, the cataloged X-ray sources from eRASS:3 are cross-correlated with brighter stars extracted from complementary optical and IR resources. Specifically, data was taken from the Gaia DR3 \citep{gdr3}, 2MASS point sources \citep{2mass} and Tycho-2 \citep{tycho2} catalogs, in addition we used the SIMBAD database. We use stars brighter than sixth magnitude in the respective photometric bands (SIMBAD/Tycho-2: B, V; Gaia: G; 2MASS: J), to obtain a stellar sample that is well suited to study optical loading in the eRASS data. As eROSITA operates in the L2 environment, Earth and Moon are no contributors in survey data. Outer planets, comets and other transient sources that might reach optically relevant brightness values are not considered here.

The characterization of optical loading not only allows the identification of 'fake' X-ray detections, but also expands the analysis possibilities for optically contaminated stellar X-ray sources. About 1500 stars are found in the V-band brightness range between 2~mag and 5~mag \citep[see e.g.][]{bsc91}; among these there are many nearby stellar systems, intrinsically bright OBA stars and giants with luminosity class III-I. While large sample studies may simply ignore all stars above a certain brightness limit, astrophysical interest in these objects may be present.

Optical loading effects on stellar sources can be quite diverse, as during survey data taking the sky is scanned in progressing and overlapping stripes and each source passes several times, but on different scan paths, through the field-of-view (FOV) of eROSITA.
If a star is sufficiently bright, a time dependent optical loading signal is generated and if enough pseudo X-ray events are created, the source detection algorithm triggers and the star makes it into the eRASS catalog, independent of its X-ray properties. For a signal to be detected, basically three different cases have to be distinguished. The star is an X-ray emitter, the star is optically bright or the star is both. While case one is described in various eRASS papers, case two and three are the focus of this work, where thresholds and relations with regard to optical brightness are studied.

If a given star is X-ray dark, or at least faint enough to fall well below the eRASS sensitivity limits, basically all the registered events are caused by optical loading. These fake X-ray sources have very soft spectra and show pseudo-variability with a dependence on the off-axis angle. If the star is optically bright and an X-ray emitter, pseudo-variability is likewise observed and its count rate and spectrum are likely be modified, whereas the relevance is governed by the contrast between the optical loading signal and the intrinsic X-ray brightness.
Any potential disentangling depends on source properties and science objective, but in general the X-ray properties of sources contaminated by optical loading are at least more uncertain.

\section{eROSITA: Survey and data analysis}
\label{obs}

This section describes the basics of the eROSITA instrument and the used software and analysis procedures.

\subsection{eRASS data analysis}

The eRASS:3 survey data has been processed with the pipeline version c030 and the eROSITA science analysis software (eSASS) as described in \cite{DR2}. The software package eSASSusers\_240410 is used to produce X-ray images, spectra and light curves. A description of the eROSITA software, source detection, catalog generation and data products is presented in \cite{esass}. The standard source detection is performed in the 0.2\,--\,2.3~keV band (single band, 1B), an energy range that is adapted to eROSITA, respectively its effective area. Aperture photometry is then used to determine the count rates in the 0.2\,--\,0.5, 0.5\,--\,1.0, 1.0\,--\,2.0 and 2.0\,--\,5.0 keV energy bands, called P1\,--\,P4. Throughout this study we adopt these energy bands and use the catalog count rates to determine hardness ratios as HR\,=\,(H\,-\,S)\,/\,(H\,+\,S), where H and S denote the respective 'hard' and 'soft' band.
To study individual stellar systems in greater detail, we use the {\tt srctool} task to analyze the event files. Source photons were extracted from circular regions around the X-ray positions, the background is typically taken from a larger annuli. If not stated otherwise, we use the merged events from all seven telescope modules as input. For images and spectra, FOV averaged properties of the eROSITA instrument apply. Count rates are, as in the eRASS catalogs, corrected for vignetting and correspond to the full instrument on-axis equivalent values.

Spectral modeling and analysis was carried out with XSPEC V12.14 \citep{xspec} in its PyXspec implementation. Absorbed multi-temperature APEC models \citep{apec}, i.e. collisionally ionized thermal plasma with solar abundances from \cite{grsa} were used to fit the X-ray spectra, optical loading was modeled with a power-law component. Spectra are background subtracted, re-binned to a minimum of three counts per bin and optimization uses the 'cstat' algorithm, that is applicable for Poisson distributed data. We tested different modeling approaches and obtained overall consistent results.
The same setup is used to simulate, using the appropriate ARF and RMF files, theoretical eRASS spectra to determine parameters like rate to flux conversions or hardness ratios. These results are then used to study the effects of optical loading, but also suited to generally characterize stellar eRASS sources.

The X-ray to optical cross-match radius is 15\,'' and coordinates are updated to epoch 2021, whenever proper motion data is available. Since the eRASS sources correspond to all emitting components within the PSF extraction radius, multiple stellar systems are tracked where information is at hand. However, optically bright systems are often dominated by a single component and in general magnitudes are not combined and differences in spatial resolution between the optical surveys are ignored. Still, for the optical flagged sample matches have the median separation of 1.7\,'' and 90\,\% are found within 6\,''. The fraction of spurious matches is checked with randomly shifted input positions and found to be negligible. On the other hand, a few sources are apparently missing due to source confusion or likely detection failures, but overall both effects mostly apply to fainter sources.

\subsection{eROSITA detectors and optical loading sensitivity}

The eROSITA instrument consists of seven identical telescope modules (TM) with CCD detectors. However, the modules are equipped with two different filter combinations and one might expect a different response to optical loading for the two types. One module type is equipped with an aluminum layer 
{with a thickness of 200 nm} on the CCD itself (TM8\,=\,TM1+2+3+4+6), while the other type has no on-chip filter, but rather a layer of half thickness added to the polyimide foil {with a thickness of also 200 nm} in the filter wheel (TM9\,=\,TM5+7). Due to the so-called light leak \citep[see][]{erosita}, i.e. light from the Sun that is unintentionally  entering  the instrument, the thresholds of TM5+7 had to be increased by about 50~eV. The sunlight does not affect the modules with the on-chip filters in a significant way. All survey data was taken in 'FILTER' mode, providing the maximum attenuation of optical light and ensuring a homogeneous dataset.

Optical loading occurs by the accumulation of low energy photons within one or several of the 9.6\,x\,9.6 arcsec sized pixel over the frame integration time of 50~ms, during which each sky-source advances by 4.5 arcsec. As the X-ray event thresholds are basically tuned to the noise of the CCDs, the light from bright sources contributes extra energy.
If the summed energy of the electronic detector noise and those of the IR-optical-UV photons from a star exceeds the set X-ray detection threshold, any sufficiently bright source starts to generate false X-ray events. These events initially appear as weak signal at lower X-ray energies and then the strength of the optical loading signal increases with increasing source brightness. For very bright sources even ‘saturation-like’ effects occur, e.g. when events become invalid and are removed by the processing chain. The optical loading signal further strongly depends on the off-axis angle of the source, because the optical PSF is sharper close to FOV center, vignetting is modest and the photons are focussed onto a smaller detector area. i.e. fewer pixels. However, this effect is largely averaged out in the survey catalogs. As optical loading adds charge, modifications in count rate and spectral distortions can also be expected for intrinsic X-ray emitter.

\begin{figure}
\includegraphics[width=90mm]{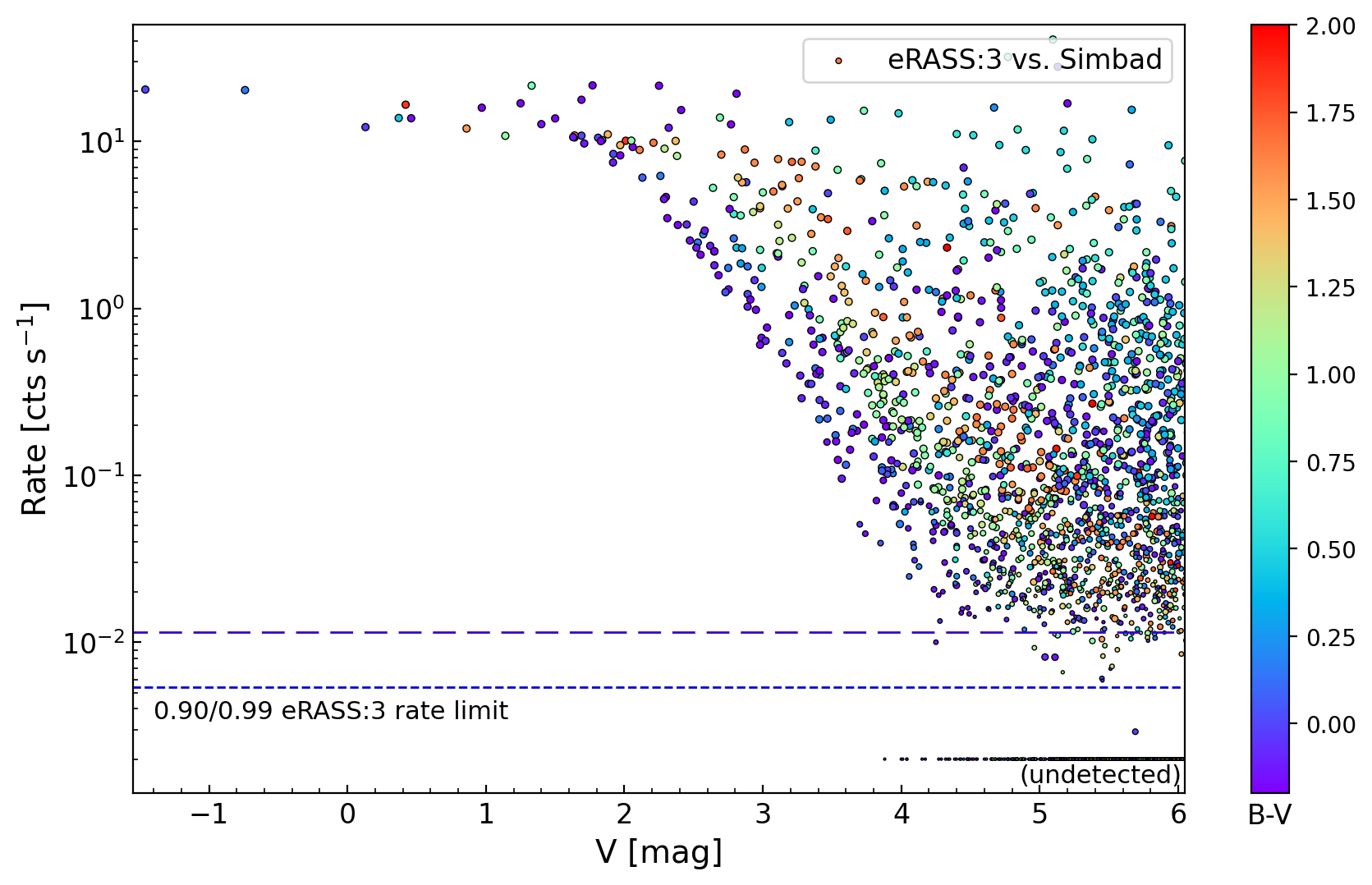}
\caption{\label{overview}The eRASS:3 catalog sources cross-matched with the SIMBAD database, X-ray count rate (1B, 0.2\,--\,2.3 keV) vs. V magnitude. The color scaling is by B-V stellar color, size increases with detection likelihood, undetected stars are plotted at $2 \times 10^{-3}$~cts\,s$^{-1}$.}
\end{figure}

\section{eRASS - optical loading effects}
\label{res_opt}

Here we report the findings obtained from our optical loading studies and their application to eRASS:3, though we expect comparable results for future releases of the eROSITA survey data.

\subsection{General characterization}
\label{opt}

To obtain an overview on optical loading, first a cross-match of the X-ray sources with bright stars {with magnitudes V $<$ 6}
from the SIMBAD database is performed. In Fig.~\ref{overview} we show the eRASS count rate vs. V~band magnitude, where three main trends can be distinguished. First, the stars brighter than about 2~mag show a largely constant count rate, second, stars between 2~mag and 5~mag show a declining minimum count rate with decreasing optical brightness, and third, beyond roughly 5~mag one enters the population, where the lower rate limit is given by the sensitivity of the X-ray data.
The overall detection fraction is 1.0 for stars brighter 4~mag, 0.91 in the 4.0\,--\,4.5 mag range, 0.80 at 4.5\,--\,5.0 mag and then drops steadily and significantly. Inspecting the few 'unexpected' stellar non-detections with $V \approx 4$~mag, one finds that they are typically located in the PSF wing of an even brighter optical or X-ray source.
The presence of optical loading sources in the X-ray catalog also depends on the exposure depth, where a deeper exposure leads to the detection of fainter stars. In eRASS these are especially the locations closer to the ecliptic poles, where a few optical loading sources in the range 5\,--\,6~mag with very low count rates are detected. On the other hand, the optical contamination of X-ray emitting stars will be mostly negligible or even non-existent at these magnitudes. An optical loading flag (\texttt{FLAG\_OPT}) is the basic output that is added to the source catalogs, as detailed for the eROSITA DR1 \citep{DR1}.
The optical loading flag in the DR2 catalogs, basing on the deeper eRASS:3 data, is set globally for sources brighter than 5.0~mag (B, V, G) and 3.5~mag (J).

\subsection{Count rates and optical catalogs}
\label{res_det}

\begin{figure}
\includegraphics[width=90mm]{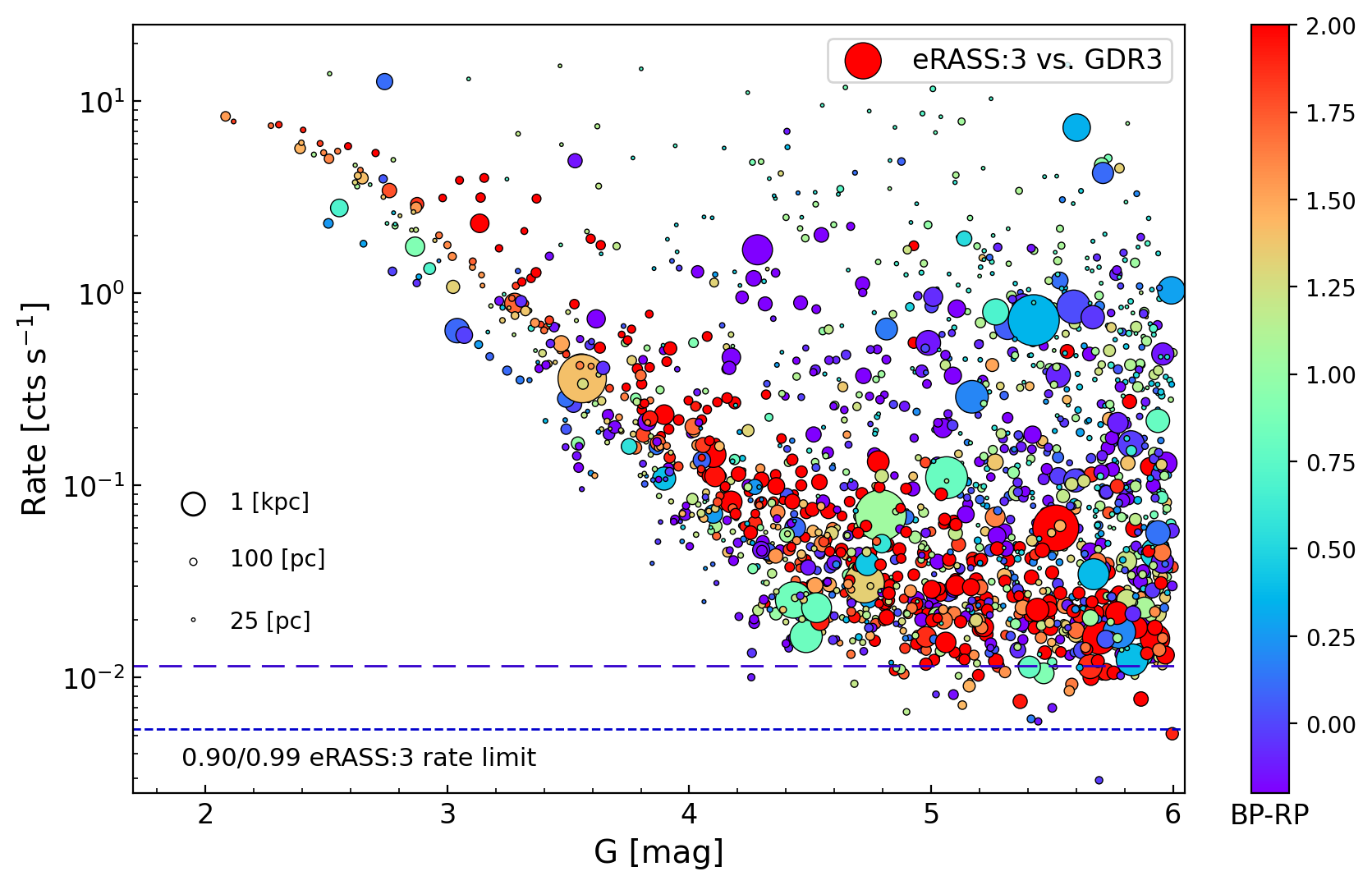}
\caption{\label{pic_gdr3_d}The eRASS:3 rate vs. G magnitude from Gaia DR3. Color is by BP-RP and size by distance, whereas all stars with $d \le 25$~pc have the smallest symbol size.}
\end{figure}

The eRASS count rate is tested against magnitudes and colors from several missions or surveys. The Tycho-2 B/V bands show comparable results to B~band taken from SIMBAD and we next analyze Gaia (G band) and 2MASS (J band); {these bands differ in  effective wavelength with $\lambda_{\rm eff} \approx  440$~nm
for the B band, $\lambda_{\rm eff} \approx 550$~nm for the V band, $\lambda_{\rm eff} \approx 670$~nm for the G band and $\lambda_{\rm eff} \approx 1235$~nm
for the J band.}
In Fig.\,\ref{pic_gdr3_d}, we plot the observed eRASS:3 X-ray count rates vs. G band magnitudes from the Gaia DR3 sample. Note that the Gaia by itself excludes very bright stars with $G \approx 2$~mag due to saturation effects in the Gaia detectors. We use Gaias superb parallax measurements to highlight the contributions of the stellar populations combined with their distance measurements. The strong contribution from giants of various colors to the bright star sample is visible.

To study optical loading in its purest form, we require effectively X-ray dark stars. For this purpose main-sequence and mildly evolved stars with spectral types mid-B to mid-A, that constitute our 'blue sample' of stars, are well suited. These stars are not magnetically active as they have no outer convective zones and their winds are too weak to generate any X-ray emission. With the caveat of known or potential binaries and a few stars that are born magnetic and where intrinsic X-rays may be expected, the remaining ones are sufficiently numerous to be suited for our optical loading studies. We further select a 'red sample', that consists of late-type giants, i.e. presumably X-ray dark stars with spectral types K/M and with luminosity class III-I.

\begin{figure}
\includegraphics[width=90mm]{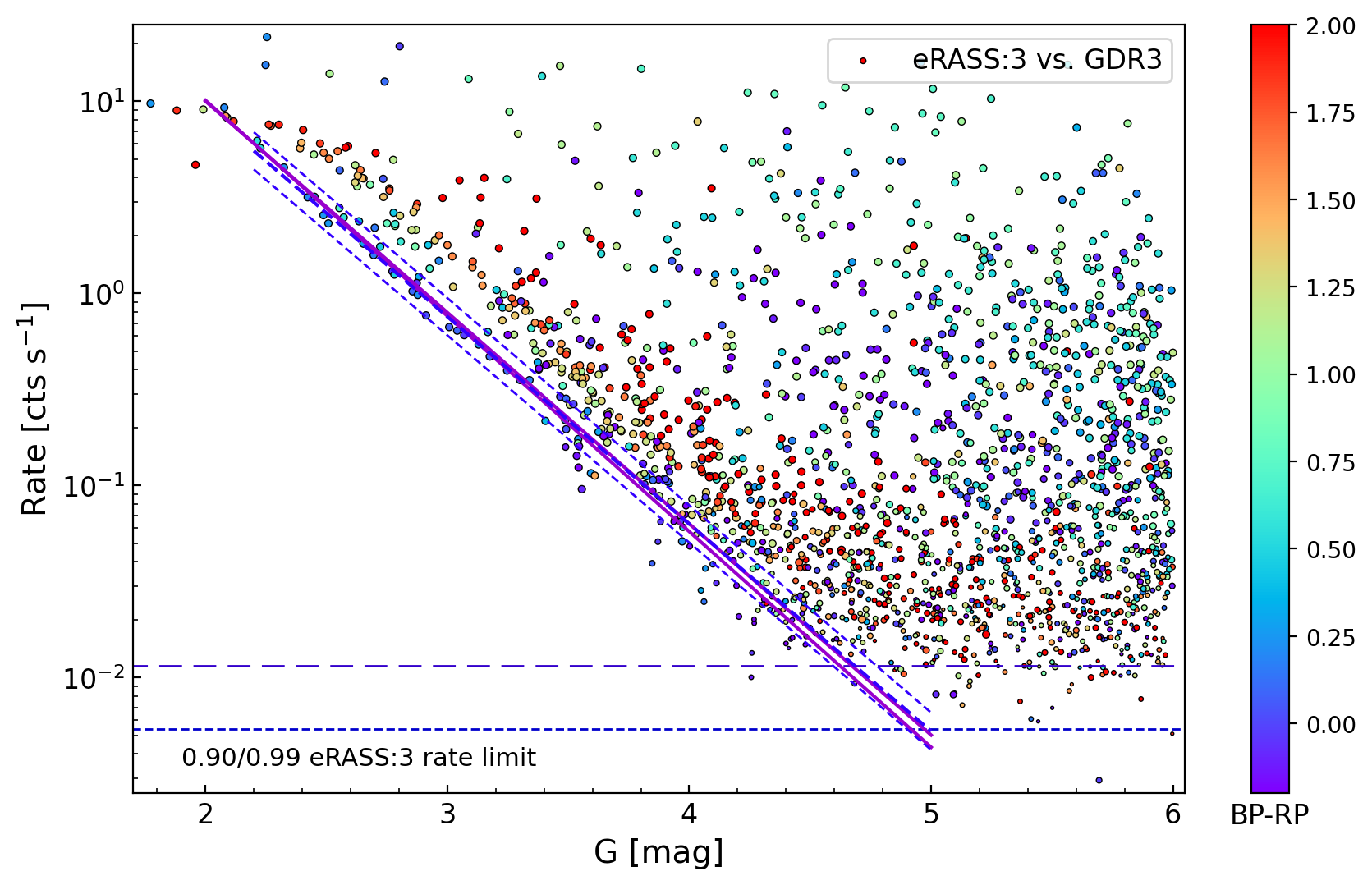}
\includegraphics[width=90mm]{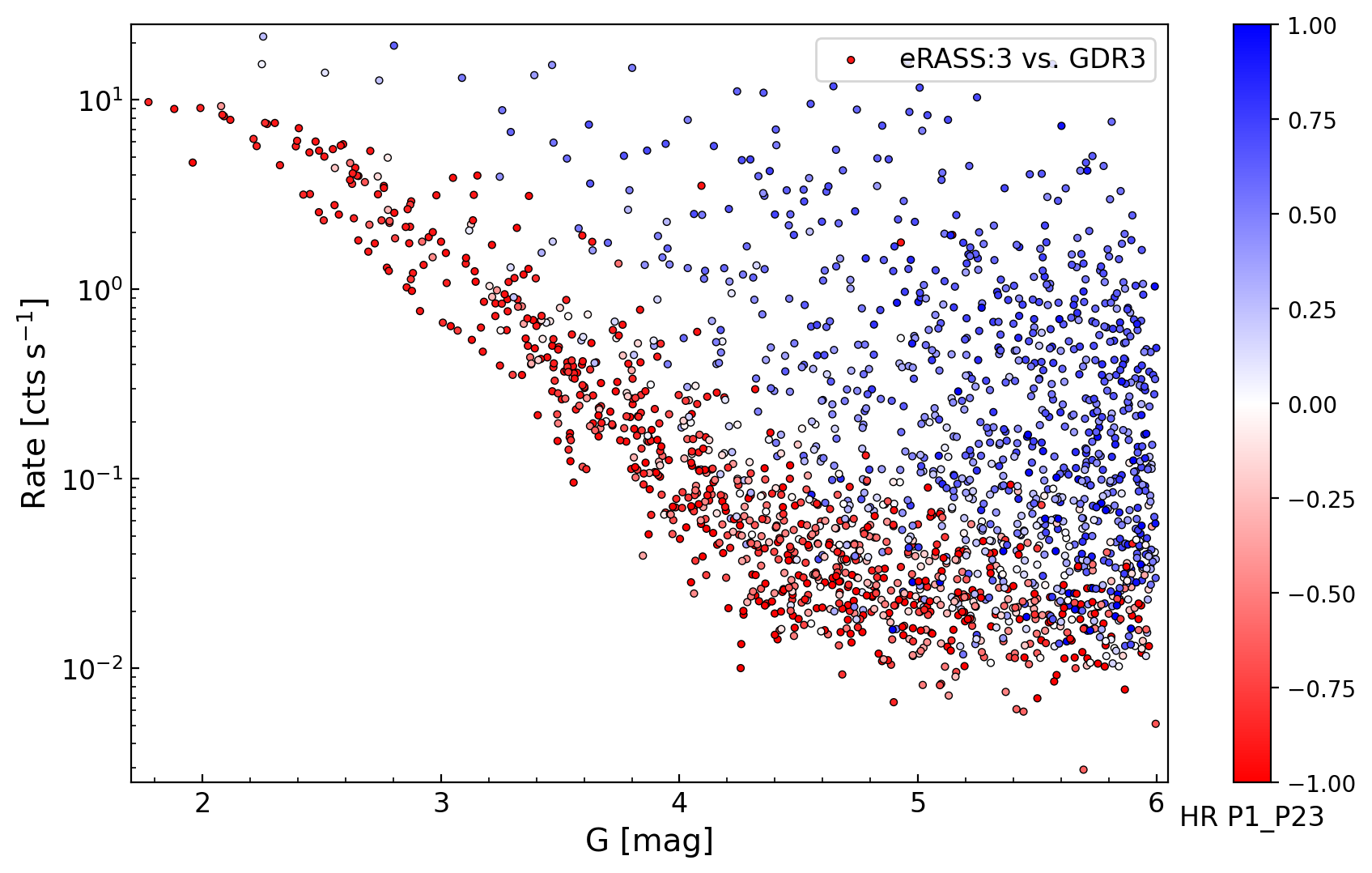}
\caption{\label{pic_rate_G}The Gaia DR3 sample. {\it Top}: Color scale by BP-RP, with the approximated optical loading rate and its $\pm$~25\,\% deviations (dashed blue) for the 'blue' stellar sample and the slope from the regression analysis (magenta). {\it Bottom}: color scaling by X-ray hardness ratio.}
\end{figure}

The eRASS rates vs. Gaia DR3 G band magnitudes are plotted in the upper panel of Fig.\ref{pic_rate_G}. While its effective wavelength with $\lambda_{\rm eff} = 640$~nm is slightly redder than the V band ($\lambda_{\rm eff} \sim 550$~nm), the trends seen in the cross-match between Gaia and eRASS data are comparable. The Gaia color BP-RP is defined by its blue (330\,--\,680~nm) and red photometer (640\,--\,1050~nm) brightness. As a further test we study the X-ray hardness ratios (lower panel), which show a very soft population gathering quite tightly around the minimum X-ray rates produced by optical loading. This population is generated by stars from our 'blue sample' with BP-RP colors (or B\,--\,V) around zero, a second population, due to intrinsic colors slightly offset, consists of stars from the 'red sample'. In addition, there is a sorting towards later spectral types and with more dominant intrinsic X-ray emission that is visible by the reddish-whitish-bluish sequence in X-ray hardness.

A quantitative inspection of the hardness ratio in P1 (0.2\,--\,0.5 keV) vs. P23 (0.5\,--\,2.0 keV) shows that stars, in which the detection is suspected to be dominated or exclusively caused by optical loading have negative values around HR $\lesssim$ -0.5. As shown in Sect.~\ref{res_cal}, these HRs would correspond to average coronal temperatures of about 1.5~MK or lower of very inactive stars that are typically X-ray faint and very rare to absent in the eRASS data.
Next are the moderately soft sources, often of later spectral types, that can be found towards optically darker or X-ray brighter stars. Typically these stars are intrinsic X-ray emitters that are also optically bright, resulting in intermediate HRs in the -0.5 to 0.5 range. A spectral hardness of HRs $\gtrsim$~0.5 is expected for coronal sources and dominates more and more in the optically fainter stars.

In the following we use the effectively X-ray dark stars from our 'blue sample' and approximate their optical loading rate for the G band magnitude vs. minimum X-ray count rate. Thereby an {empirical} analytical description was determined and a regression analysis of the eRASS:3 count rates vs. the G band magnitudes between 2.25\,--\,5.0 mag was performed, where we use rate quantiles of 0.25 and 0.3 per brightness bin. The best-fit relation between minimum X-ray rate and stellar brightness is:

\begin{equation}
\label{eq_opt}
\mathrm{Rate}_{\mathrm{Xopt}} = 0.00525 \times 12.0^{~(5-\mathrm{G})}~~\mathrm{[cts/s]}
\end{equation}
with the stellar brightness measured in G magnitude.

As shown by the red and blue lines in Fig.~\ref{pic_rate_G}, {the ansatz described by Eq.~\ref{eq_opt} describes the observational results very well.}
However, especially at the fainter end observational properties like different individual scan paths and noise effects start to play a more important role. The total modeled rate can be split into the contributions from the photometric bands, where we find that about 95\,\% is emitted in the 0.2\,--\,0.5~keV (P1) band and the remaining at 0.5\,--\,1.0~keV (P2) energies. A median spectral hardness of HR\,=\,-0.85 for the sample provides further support that optical loading is basically the sole contributor in the selected stars.

\begin{figure}
\includegraphics[width=90mm]{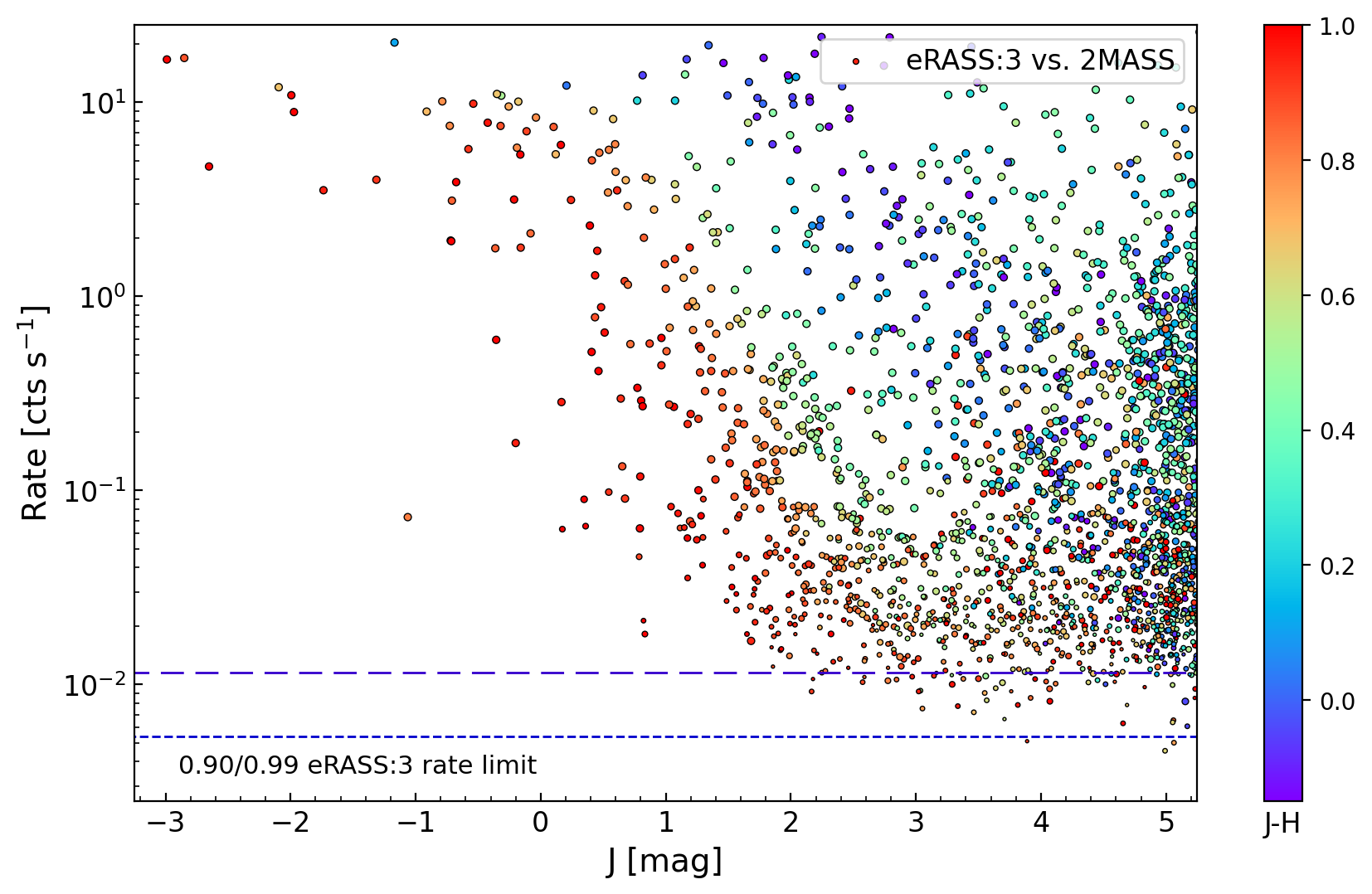}
\includegraphics[width=90mm]{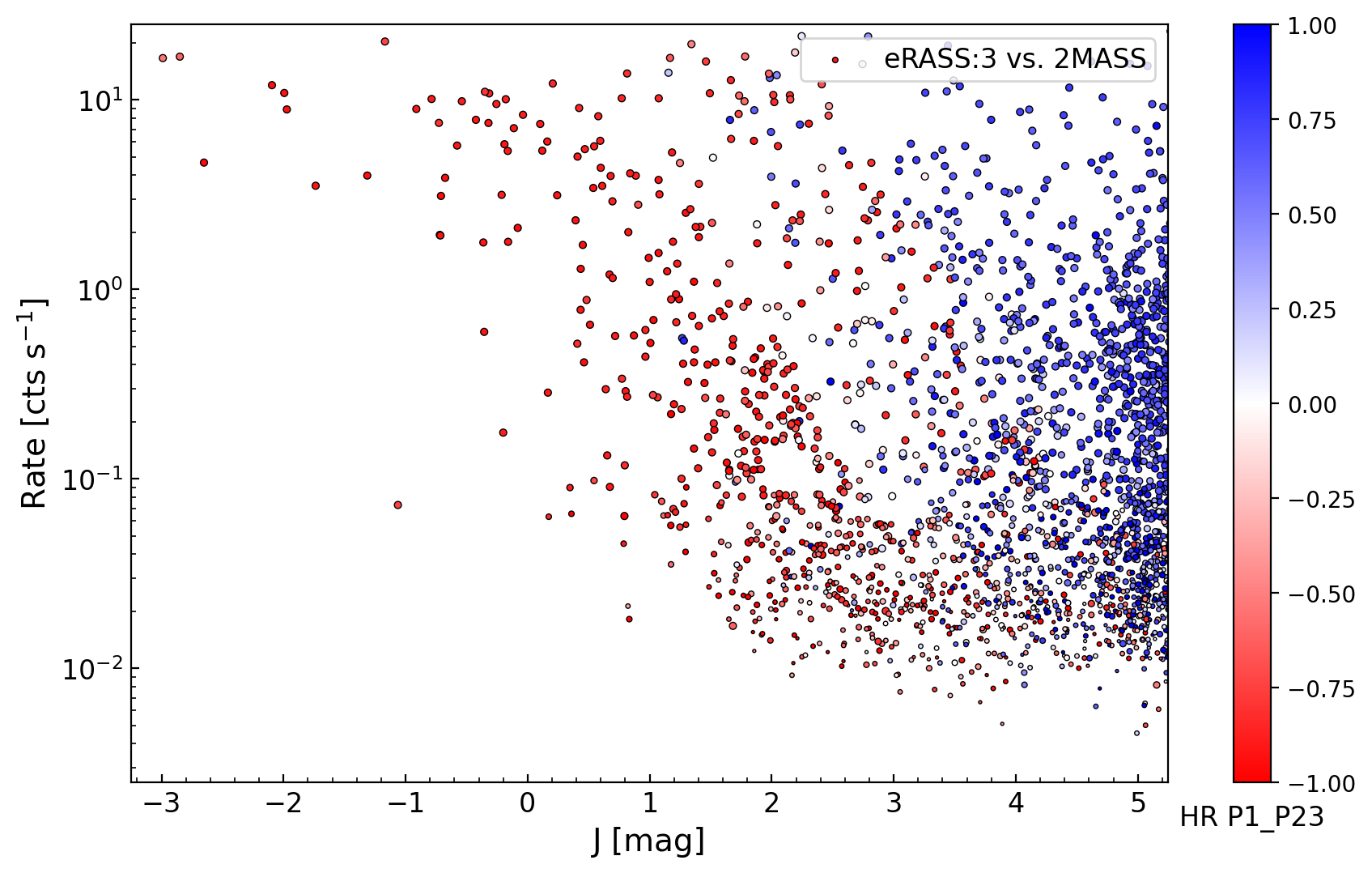}
\caption{\label{pic_rate_J}The 2MASS sample; {\it Top}: color scale by J-H and {\it bottom}: X-ray hardness ratio.}
\end{figure}

For main-sequence stars with a sufficiently large count rate and spectral hardness, the presence of intrinsic stellar X-ray emission is very likely. Likewise, the effects on the observed rate are moderate, if there is an adequate contrast between X-ray emission and the optical loading signal. For these cases the X-ray data is usable with some limitations. If the corrected rate, i.e. observed rate minus optical rate, is at least above the optical rate multiplied by some factor and additionally exceeds a minimum rate, a 'true' X-ray detection is very likely. {Various experiments with the data show that} a multiplication factor of three and a minimum rate of 0.05 cts/s are practical estimates for stars in a brightness range of 2.5\,--\,5~mag, the used multiplication factor and minimum rate do of course depend on the required accuracy and the science case.

The correlation of eRASS count rates to the NIR brightness, here J~mag as measured by 2MASS (1.235\,$\mu$m), is shown in Fig.~\ref{pic_rate_J}. The spectral types are again well separated, but the bright end is quite fuzzy and the slope of a minimum count rate distribution is less well defined. Among the sources that are particular bright are many red giants, which are known to be significantly variable.
As with the G band data, we find that the present outliers are mostly due to stellar variability as the data is not taken simultaneously. Notable causes are e.g. pulsations, occultations and rotational modulation and stars in multiple systems or dusty environments.
The HR plot (Fig.~\ref{pic_rate_J}, bottom) highlights the soft spectra of the 'red sample' stars and also shows the softness of the stars from the 'blue sample' around color zero. Still, there are quite a few bright sources that are surprisingly hard in X-rays. Many of these are late-type giants, but again binarity is an issue and on inspection these turn out to be known or likely symbiotic stars.

While NIR brightness has to be considered, we refrained from a modeling as was done for the G band due to the large scatter and estimate a limit from the count rates and hardness ratios of the detected sources. The optical loading flag in the eRASS catalog is present for sources brighter than J\,=\,3.5~mag.

\subsection{eRASS vs. RASS}
\label{res_rass}

\begin{figure}
\includegraphics[width=90mm]{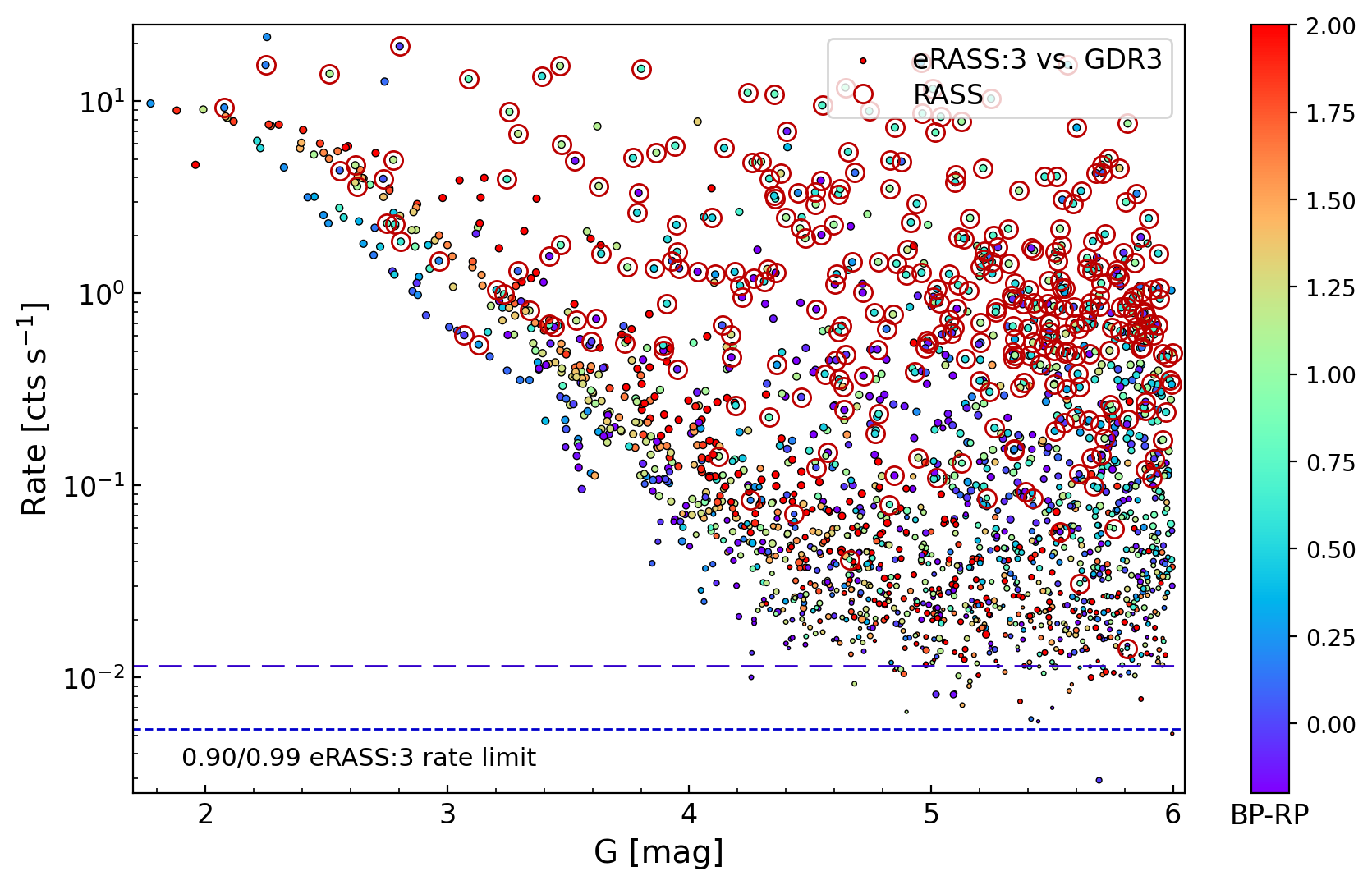}
\caption{\label{pic_rass}eRASS:3 rate vs. G magnitude as in Fig.~\ref{pic_rate_G}, the RASS detected stars 
are overplotted as red circles; {the dashed and dotted lines indicate those rates that 90\%/99\% of the 
detected sources exceed.}}
\end{figure}

The final test is made by comparing different X-ray measurements for the same source. Here we use the RASS in its 2RXS version \citep{2RXS}, adopt a minimum likelihood of eight and repeat the matching procedure as before on eRASS-Gaia detected stars. The RASS is shallower than the eRASS, but here the detected sources are without any contribution from optical contamination.

The result is shown in Fig.~\ref{pic_rass} and in the regime of X-ray brighter coronal emitter virtually all stars are recovered, the few missing ones are likely due to variability, positional accuracy etc. However, basically none of the presumed pure optical loading stars, i.e. those near the minimum rate slope shown in Fig.~\ref{pic_rate_G}, has a RASS detection and is unlikely to be a 'true' X-ray source. There a two exceptions, $\mu$.01 Sco (B1V+B) and $\lambda$~Cen (B9III), both located in the Sco-Cen association and known/likely multiple systems. A mixed detection fraction is present for early-type giants, but is in late-type giants close to zero as expected for optical loading stars. Independent of stellar brightness, the RASS detection fraction fades out at lower eRASS count rates due to sensitivity issues.

Very similar trends are seen by a comparison to the V~mag data from SIMBAD. Here one finds at the very bright end the binary system Sirius (A0+WD), where the RASS detects X-rays from Sirius~B, while the eRASS data is strongly contaminated by optical light from Sirius~A.

\begin{figure}
\raisebox{0.19cm}{\includegraphics[width=56mm]{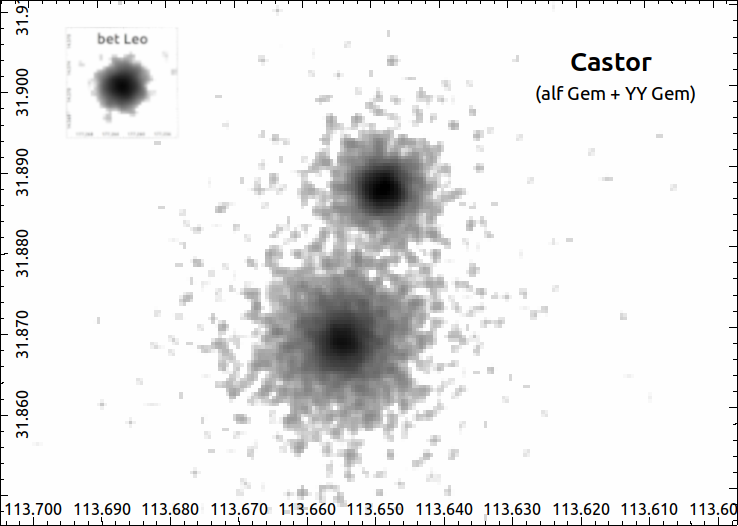}}
\hspace{0.04cm}
\includegraphics[width=32mm]{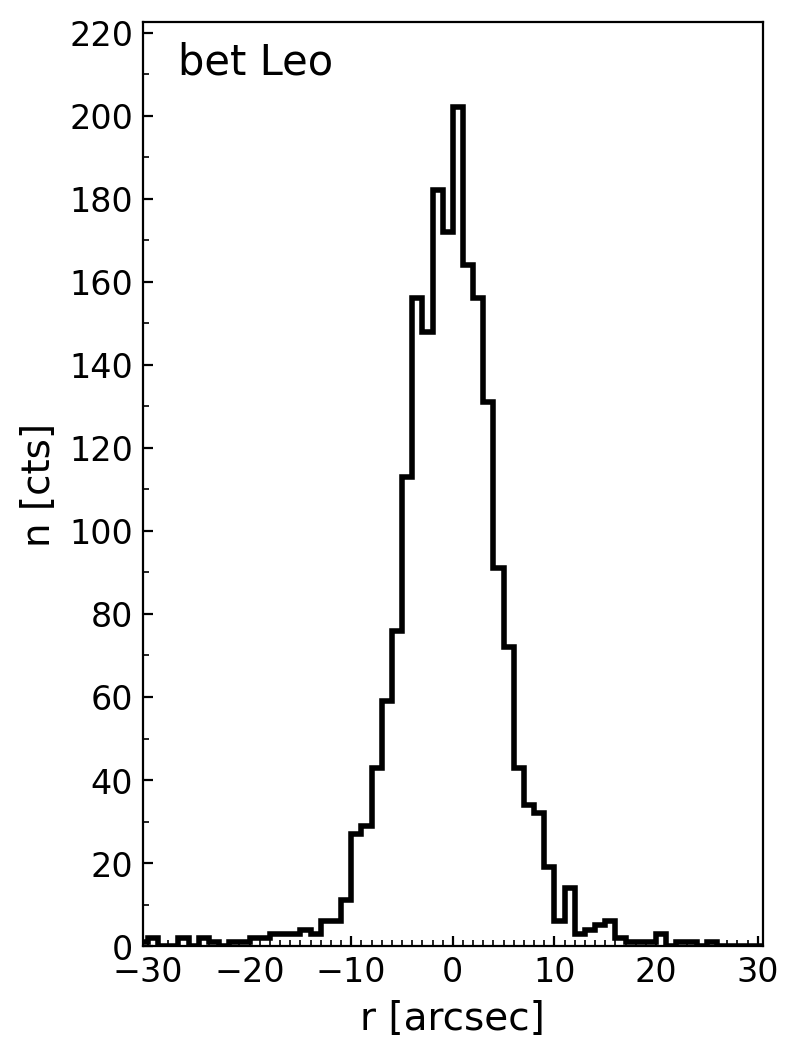}
\caption{\label{pic}eRASS image of the Castor system in equ. coordinates, depicting an X-ray plus optically bright northern component ($\alpha$~Gem) and a solely X-ray bright southern component (YY~Gem). A exemplary, pure optical loading source ($\beta$~Leo) is inserted. The images are logarithmic, gaussian smoothed and on the same spatial scale; their exposure times are comparable. A 1D-projection of $\beta$~Leo is shown on the right.}
\end{figure}

\subsection{X-ray images and PSF shape}
\label{img}

As an example of optical loading in eRASS image data, we show the Castor system in Fig.~\ref{pic}. Actually Castor, located at a distance of 15~pc, is a complex stellar system containing three known binaries, Castor A+B (V\,=\,1.6~mag) with spectral types A1V+A2V and Castor C (V\,=\,9.3~mag) with spectral type M\,0.5. In eRASS only two resolved sources are visible, whereas the northern component Castor A/B aka $\alpha$~Gem is optically bright and an X-ray emitter, while the southern component Castor~C aka YY~Gem is only X-ray bright. Their contrast ratio in the optical V-band is about 1000, whereas the eRASS rates are similar. While the early A-stars are presumably X-ray dark, Castor A and B are itself binaries as indicated by spatially resolved data from XMM-Newton/Chandra \citep{stelzer03}, each with a late-type companion. Nevertheless, in eRASS data the detected events from Castor A/B are predominantly created by optical loading, while those from Castor~C are not. This is reflected in the different average PSF width of these sources.

\begin{figure}
\hspace{0.87cm}\includegraphics[width=74.6mm]{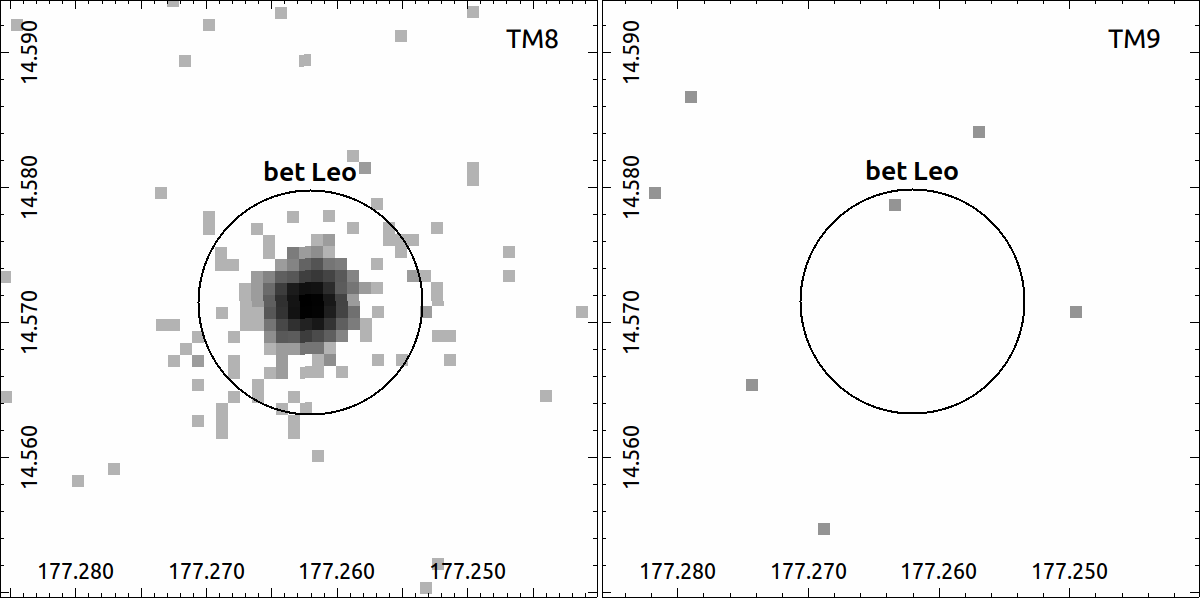}

\hspace{0.3cm}\includegraphics[width=40mm]{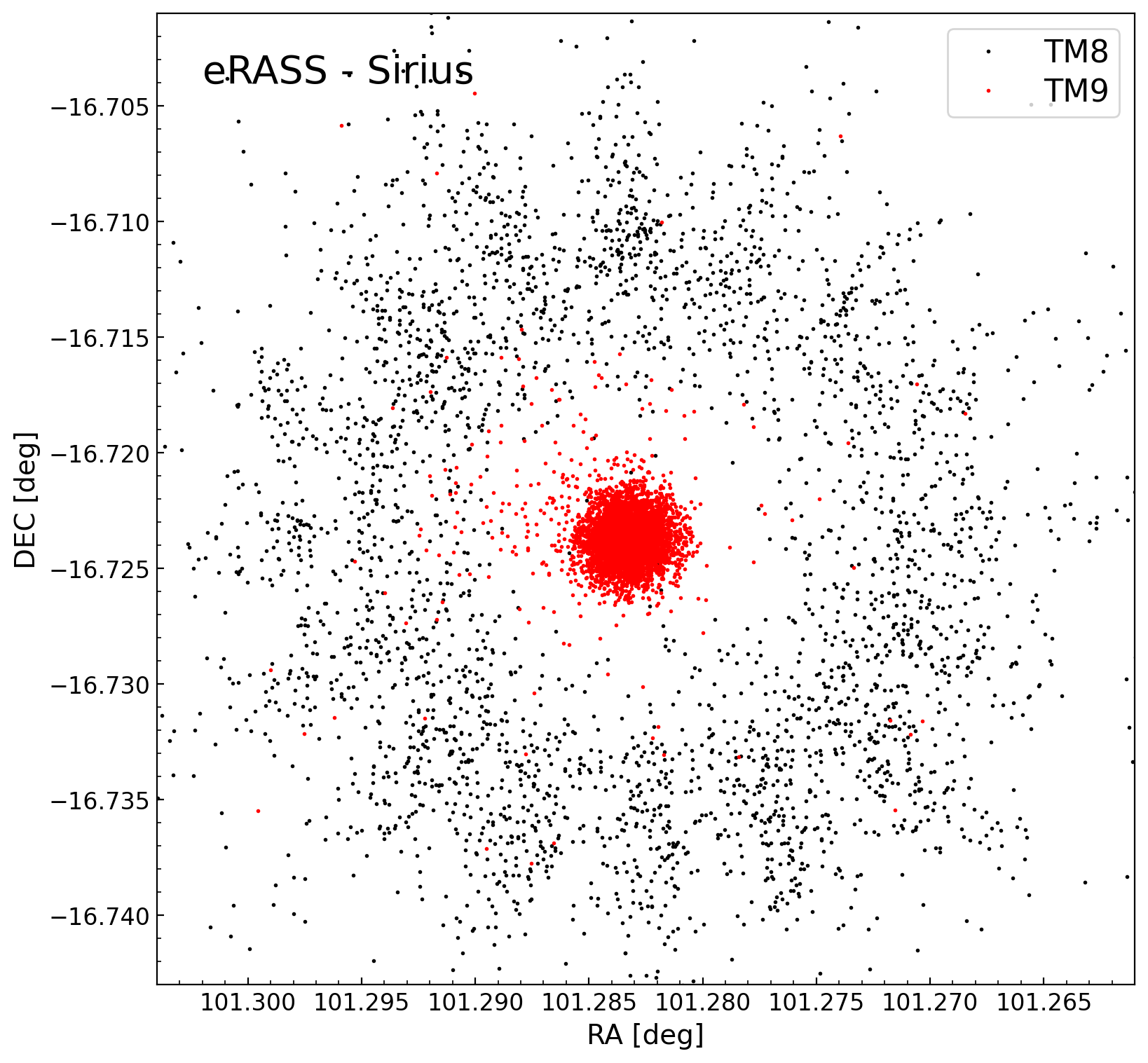}
\includegraphics[width=40mm]{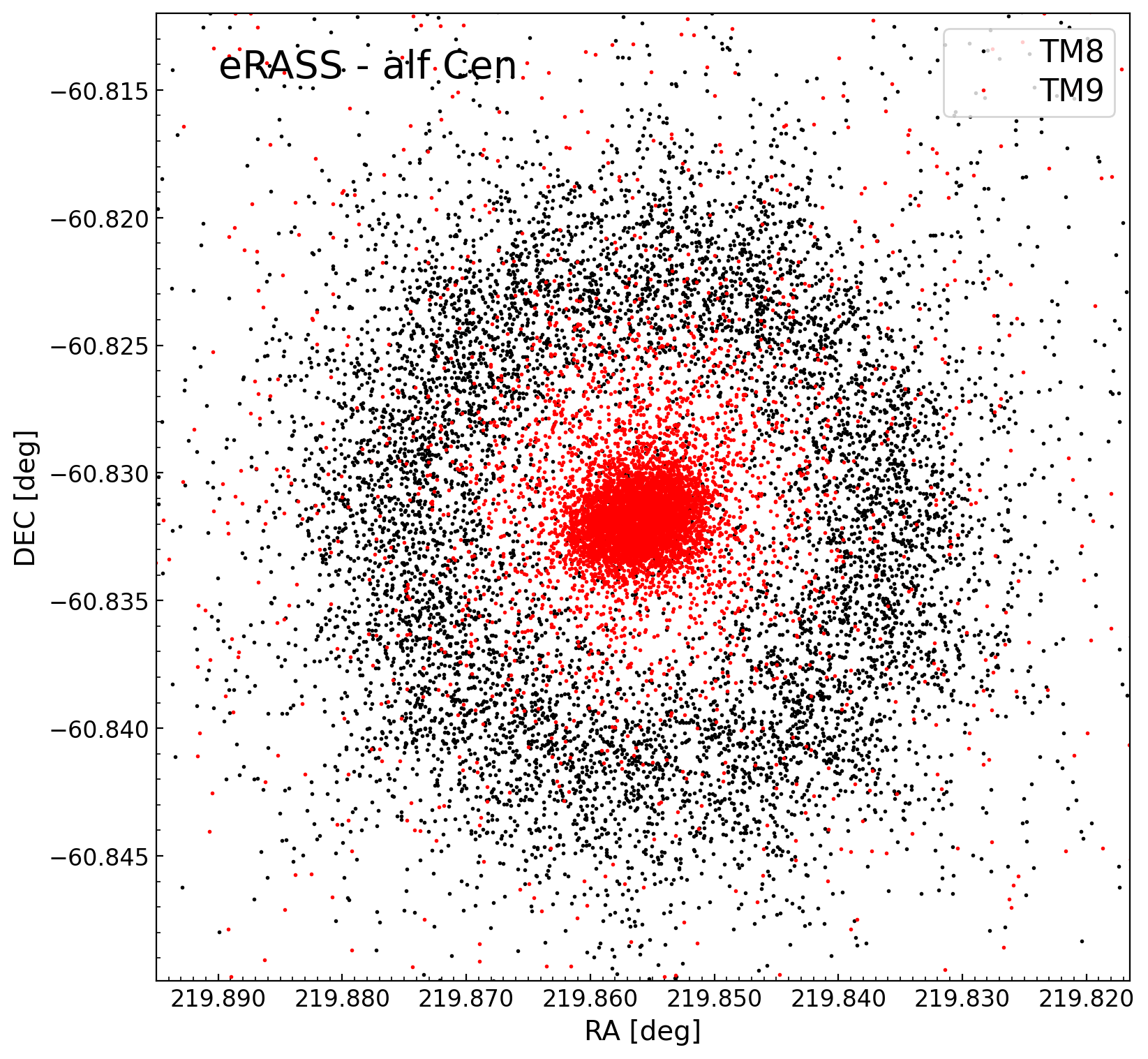}
\caption{\label{pic_opttm}Images subdivided by TM type, {\it top}: the pure optical loading source $\beta$~Leo (V = 2.1 mag), {\it bottom}: the X-ray plus optical bright sources Sirius (V = -1.5 mag) and $\alpha$~Cen A+B (V = -0.1 mag).}
\end{figure}

The optical PSF misses the distinct wings of its X-ray pendant, as shown by the pure optical source $\beta$~Leo (A3V, V\,=\,2.1~mag) as inset in Fig.~\ref{pic} and in the right panel as 1D-projection of the events up to 2~keV. Its FWHM is about 10\,'', compared to 17\,'' for X-ray sources, and basically all events are found within a radius of 10\,''. This shows that only very few pixel around the position of the stars are illuminated at levels that exceed the trigger threshold.
In addition, as detailed in Sect.\,\ref{res_spec}, the optical loading events are exclusively from the telescope modules with on-chip filter (TM8). This also affects the standard images that are generated from all modules as shown in Fig.~\ref{pic_opttm}. For moderately bright, pure optical loading sources, TM8 shows a point-like source, while TM9 is blank. If the source is also an X-ray emitter, an overlay of the X-ray plus loading signal from TM8 and the X-ray signal from TM9 occurs. Depending on the brightness and contrast, this generates point-like to ring-like structures from the TM8 photons plus the X-ray source from TM9. 
The {lack  of the 'inner' events caused by optical loading} in TM8 is clearly visible.

While these images illustrate the focusing properties and event generation of optical or X-ray photons respectively in an integrated fashion, the off-axis angle dependence of the strength of the optical signal is better visible in time-resolved data.

\subsection{Light curves and pseudo variability}
\label{res_var}

Pseudo-variability is another effect of optical loading, as the PSF and thereby the spatial extent of the optical charge cloud is strongly off-axis dependent. Variability generated by optical loading is thus present in the intra-scan count rates, i.e. light curves on timescales of seconds during a FOV passage, as well as the scan-to-scan averaged rates and hence in light curves on timescales of hours or days. Here we refer to already vignetting corrected light curves.

\begin{figure}
\includegraphics[width=89mm]{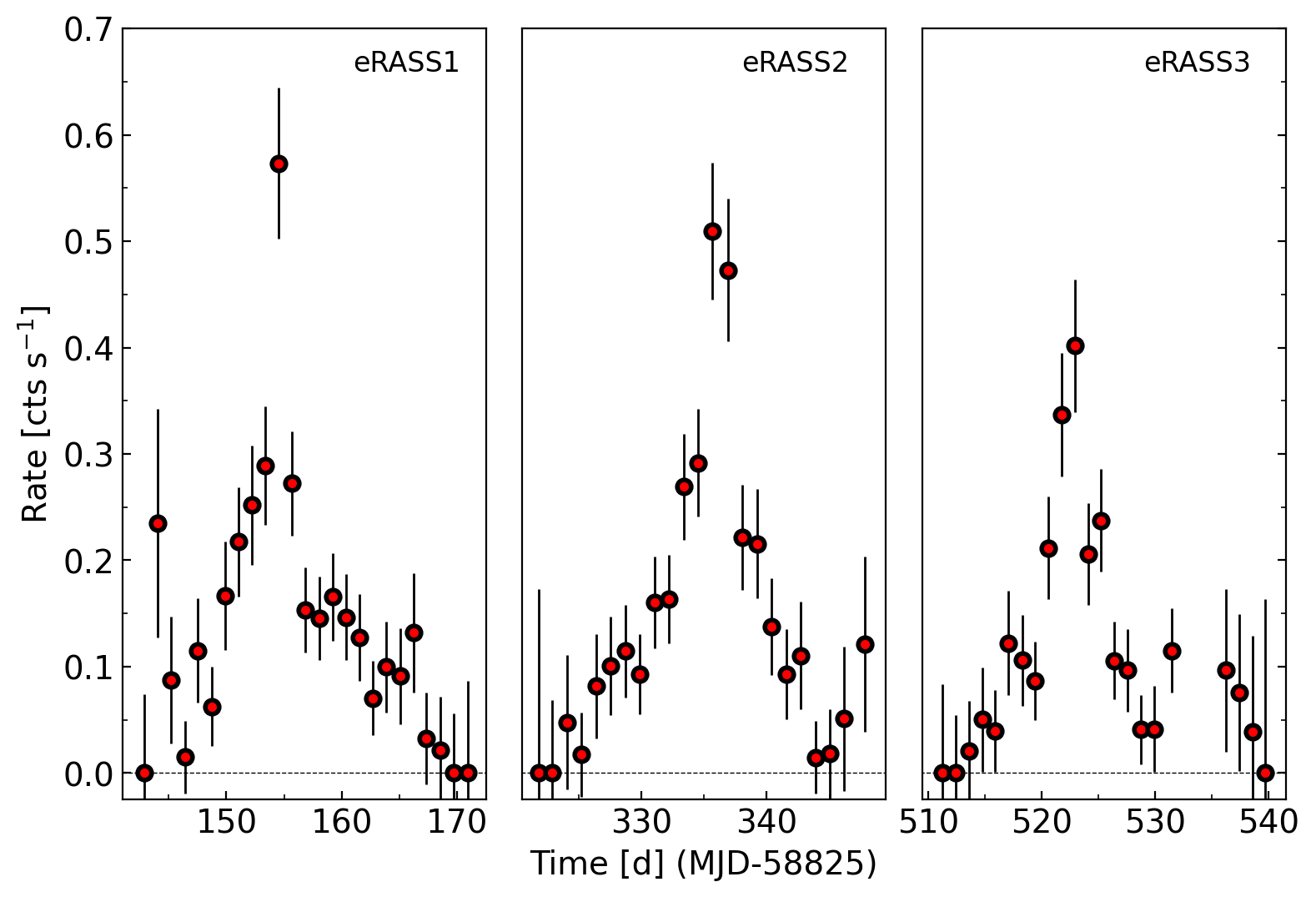}
\caption{\label{var}The eRASS:3 pseudo-variability of an optical loading generated source, here the giant M2.5III star $\eta$\,02~Dor, binning is roughly 1~day.}
\end{figure}

An example that highlights the effects on long-term light curves is given in Fig.~\ref{var}. The scan-based variability refers to count rates averaged over one or multiple scans, here shown is the eRASS rate with roughly daily time bins for the M2.5III giant HR~2245 (J\,=\,1.6, G\,=\,4.1 mag), aka $\eta$\,02~Dor as observed over the first three surveys. The star is located close to the southern ecliptic pole at latitude of -88.5~deg and has a coverage of about 30~days in each 0.5~yr eRASS survey. In each eRASS a roughly triangular pattern re-appears in the daily averaged count rate, that reflects the underlying survey observing pattern, while individual scan-paths pass the FOV-center at different impact distances.

\begin{figure}
\includegraphics[width=88mm]{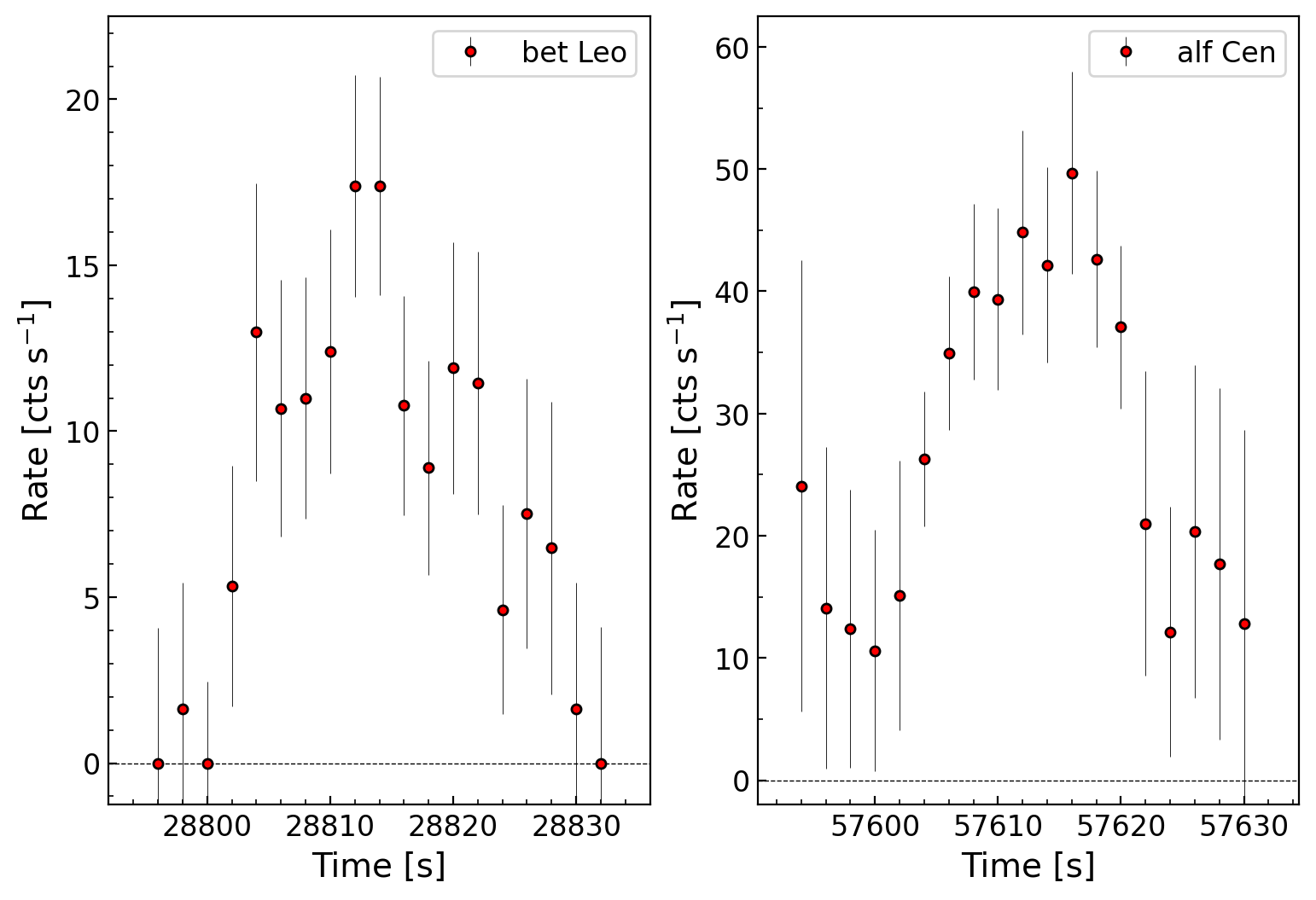}
\caption{\label{var2}A central FOV scan during eRASS1 of $\beta$~Leo (left) and $\alpha$~Cen A+B (right),
 binning is 2~s and time-zero the start of source coverage.}
\end{figure}

Short-term features such as the intra-scan variability become apparent over timescales of seconds in highly time-resolved light curves of the brighter sources. It occurs during the up to 40~s coverage of the source during the passage through the eROSITA FOV and the main feature is again an apparent brightening followed by a dimming due to the off-axis dependent optical PSF. As an example the light-curves from a central scan of $\beta$~Leo (A3V, V\,=\,2.1 mag) and $\alpha$~Cen (G2V+K1V, V\,=\,-0.1 mag) are shown in Fig.~\ref{var2}. The star $\beta$~Leo is a pure optical loading source with the typical triangular shaped light curve that basically declines to zero in the outer FOV area. Here the charge cloud from the optical light is distributed over a larger region and thereby many pixels, thus preventing triggering. As the events are mostly from the inner FOV region, which has a narrower PSF, the sources appear also sharper in the images. In contrast, $\alpha$~Cen is also an intrinsic X-ray emitter that generates an additional positive baseline level in count rate and a broader overall PSF. The $\alpha$~Cen data is not suitable to accurately determine its basic X-ray properties due to the strong corruption of the X-ray signal by optical loading, but illustrative for this type of sources due to its high flux at X-ray energies and in the optical regime.

\subsection{Spectra and spectral distortion}
\label{res_spec}

As optical loading is caused by the accumulation of many low-energy photons that add to the omnipresent detector noise, the strongest effects are present at the softest X-ray energies. It is found in a spectral modeling of the created distribution of pseudo X-rays, that these can be overall described by a power-law component. It has a rather uniform slope and a normalization given by the source brightness.

\begin{figure}
\includegraphics[width=88mm]{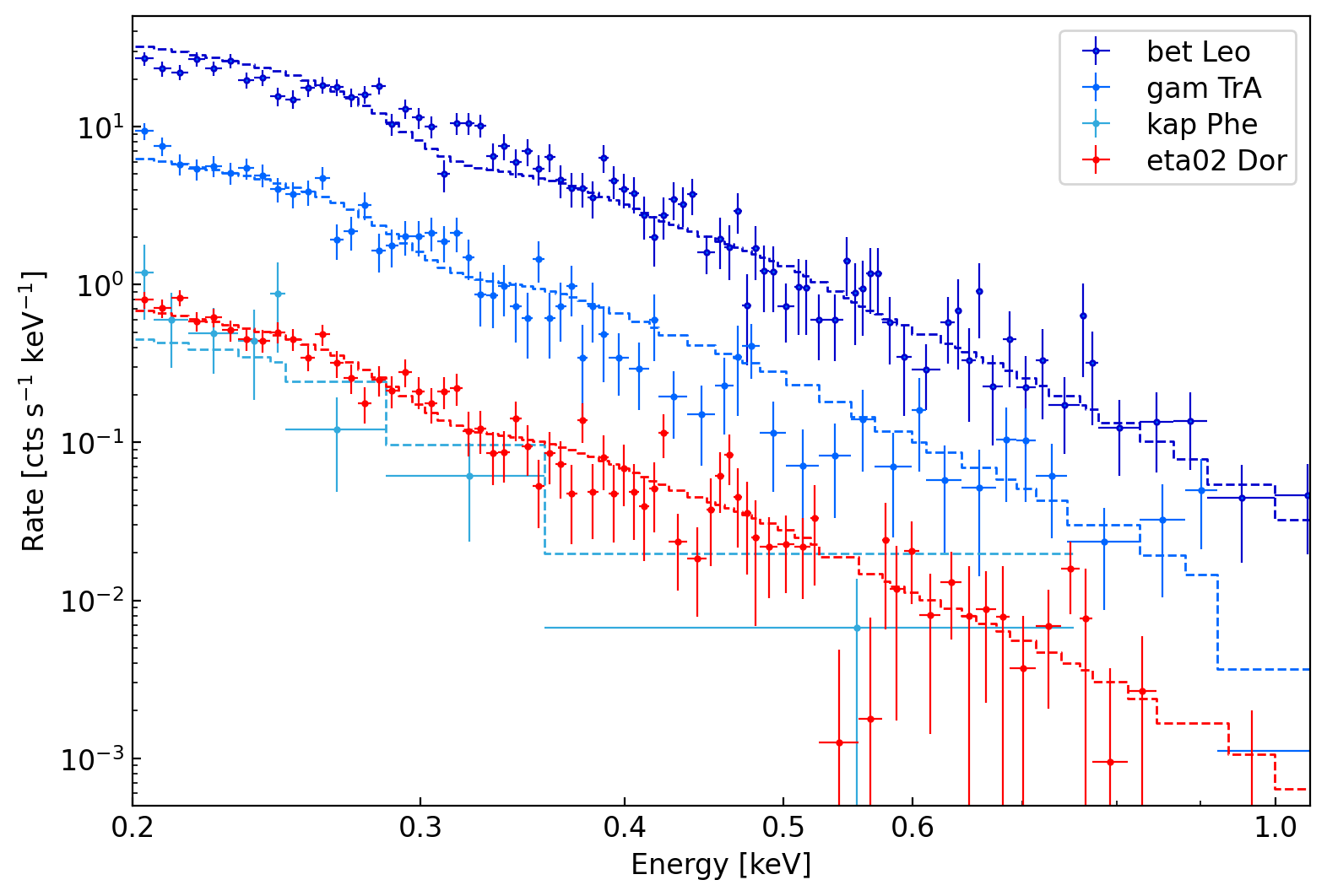}
\caption{\label{spec_opt}The eRASS spectra of four optical loading stars, each modeled with a power-law of identical slope.}
\end{figure}

Four exemplary eRASS spectra of optical loading sources are shown in Fig.~\ref{spec_opt}, with the spectral quality depending on the respective exposure time, i.e. sky location. These stars have different optical brightnesses and spectral types, but all are thought to have no intrinsic X-ray emission, i.e. the early A-type stars $\beta$~Leo (A3V, V\,=\,2.13, J\,=\,1.85), $\gamma$~TrA (A1V, V\,=\,2.89, J\,=\,2.84), $\kappa$~Phe (A5IV, V\,=\,3.94, J\,=\,3.63) and the red giant $\eta$\,02~Dor (M2.5III, V\,=\,5.01, J\,=\,1.63).

The optical loading stars are modeled simultaneously with a single power-law, that has a free but tied exponent and a free normalization for each star. We find that a slope of $\alpha = 6.1 \pm 0.1$ describes our data quite well. The derived optically induced flux depends a bit on the exact energy range considered (here 0.2\,--\,2.0~keV), the spectral binning or the used fitting method, errors are in the range $\pm 20$~\%, but mostly typically lower. Similarly, the fluxes are also within the same error range, if each star that has sufficient signal is fitted with an individual power-law exponent.
Fig.~\ref{spec_opt} shows the spectrum and model for the mentioned stars and fitted with a universal power-law component with an exponent of $\alpha = 6.1$. The theoretical HR\_P1\_P23 hardness ratio of this model is $HR=-0.9 \pm 0.02$ and robust against reasonable variations. This is validated for the power-law exponent ($\alpha =5.8\dots 6.4$) and including moderate absorption (${\rm NH} \lesssim 10^{20}$\,cm\,$^{-2}$), realistic for the analyzed stars. These hardness values are by far not reached in typical coronal plasma models and would require a very low temperature of 1~MK or below. Stars with these coronal temperatures do exist, but as their $L_{\rm X}$ is very low, they are virtually not detected by the eRASS and if they would be very nearby and thus optically quite bright. The individual spectra are influenced by the distribution of the optical light on the traversed scan paths and stellar optical properties, but overall the uniform power-law model is a quite good description. The findings from the spectral modeling agree very well with the hardness ratios found in real data as described in Sect.~\ref{res_det} and with the simulated data discussed in Sect.~\ref{res_cal}.

\begin{figure}
\includegraphics[width=88mm]{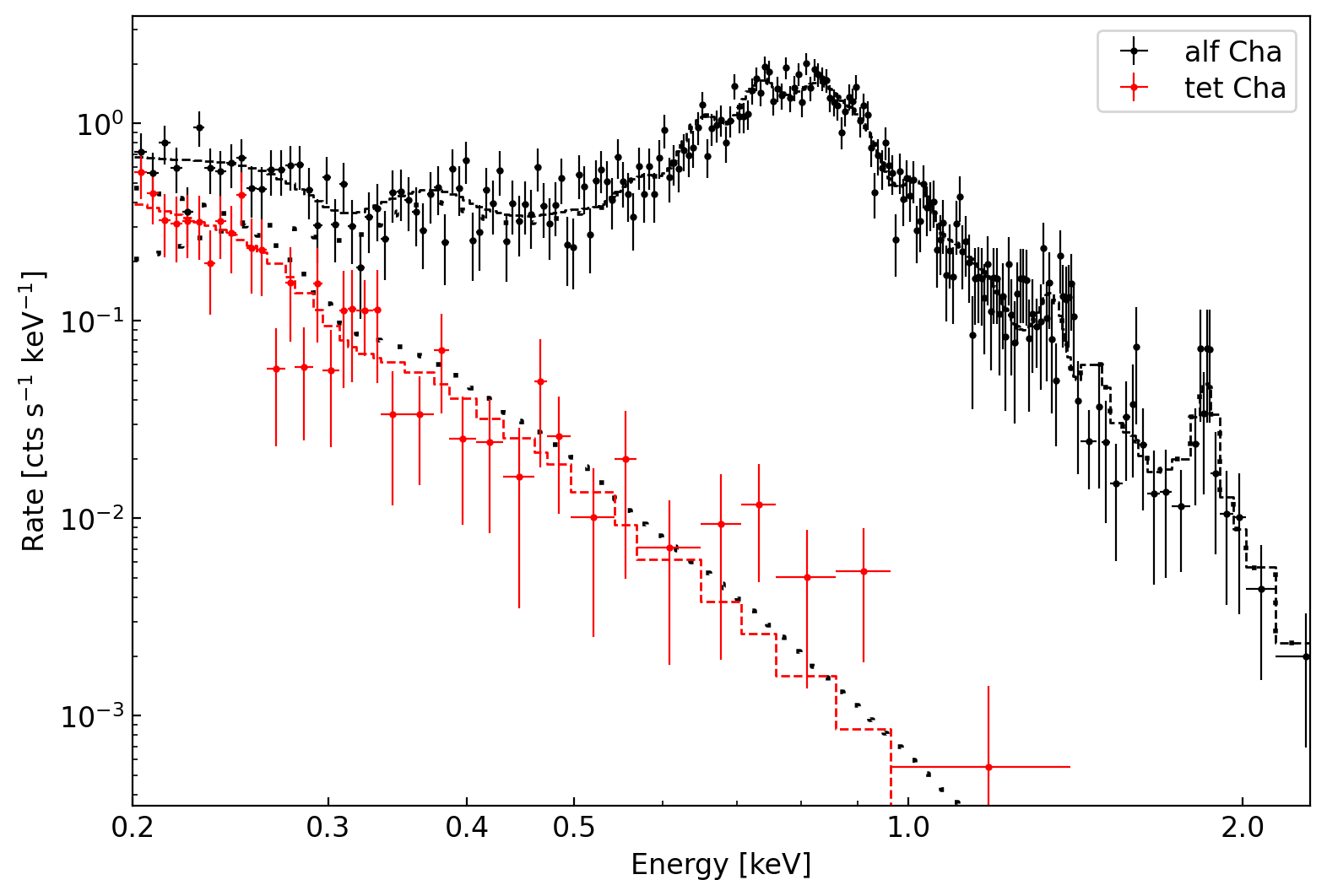}
\caption{\label{spec_xopt}Two similar optically bright stars with opt. loading and X-rays (black) and optical loading only (red); spectral models (dashed) and $\alpha$~Cha components (dotted) are plotted.}
\end{figure}

Spectral distortion requires, that the star is in also an emitter of real X-ray photons and both signals overlap. A comparison between the optical\,+\,X-ray source $\alpha$~Cha (F5V, V\,=\,4.05, J\,=\,3.42) and the optical source $\theta$~Cha (K2III, V\,=\,4.34, J\,=\,2.24) is shown in Fig.~\ref{spec_xopt}. The spectrum of $\alpha$~Cha is fitted with an thermal plasma model plus a power-law component for the optical contribution, $\theta$~Cha only with the optical power-law. The summed model describes the total spectrum of $\alpha$~Cha very well and one finds that about 90\,\% of the observed flux is from the X-ray component. The strength of the optical component is nearly identical for both stars, as shown in the comparison of the respective optical loading component.

For a star like $\alpha$~Cha, a fit with a pure plasma model but applied to the spectrum above 0.3~keV would also do quite well. It gives similar results for the basic coronal properties, as the relative contribution from optical emission declines quite fast at higher energies. Of course the energy cut value depends on the brightness and required accuracy, as it just ignores the contribution from the optical light. Any potential spectral distortion is found here to be moderate and spectral lines are roughly at the expected energies, given the available spectral resolution.

If spectral distortion is of concern, depends on the stellar optical brightness and also on the respective scientific objective. However, due to the present off-angle dependence of optical loading, the added charge is variable and might become virtually absent, especially for optically fainter stars. For relatively small contamination, the optical loading events can be quite well included as an additional component in the spectral models, as for example applied to the case of $\eta$~Car in \cite{2024A&A...682A.172S}.

\begin{figure}
\includegraphics[width=88mm]{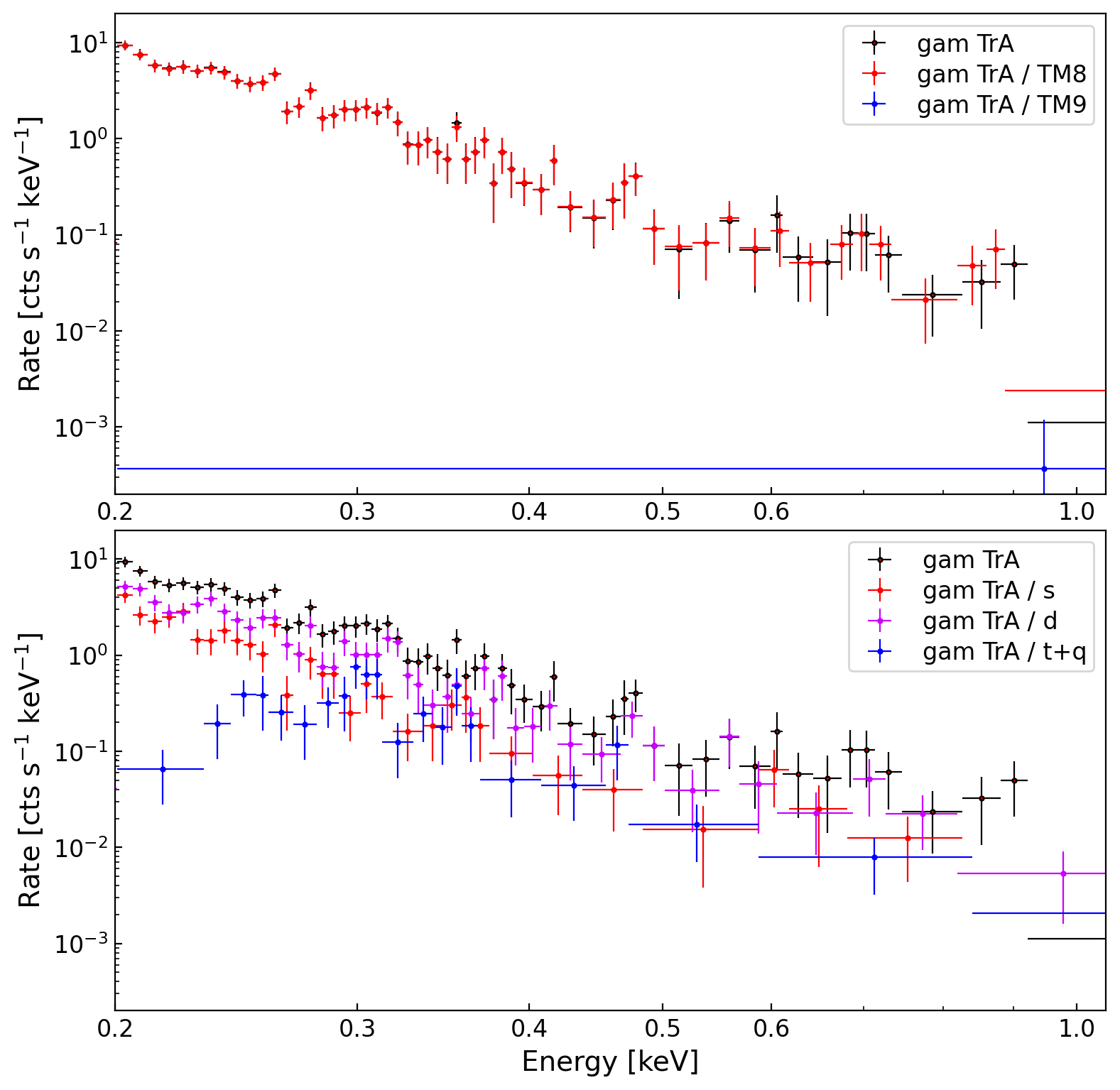}
\caption{\label{tmpat}The spectrum of $\gamma$~TrA with V\,=\,2.9~mag and the contributions per TM type (top) and event pattern (bottom); {see text for details}.}
\end{figure}

\begin{figure}
\includegraphics[width=88mm]{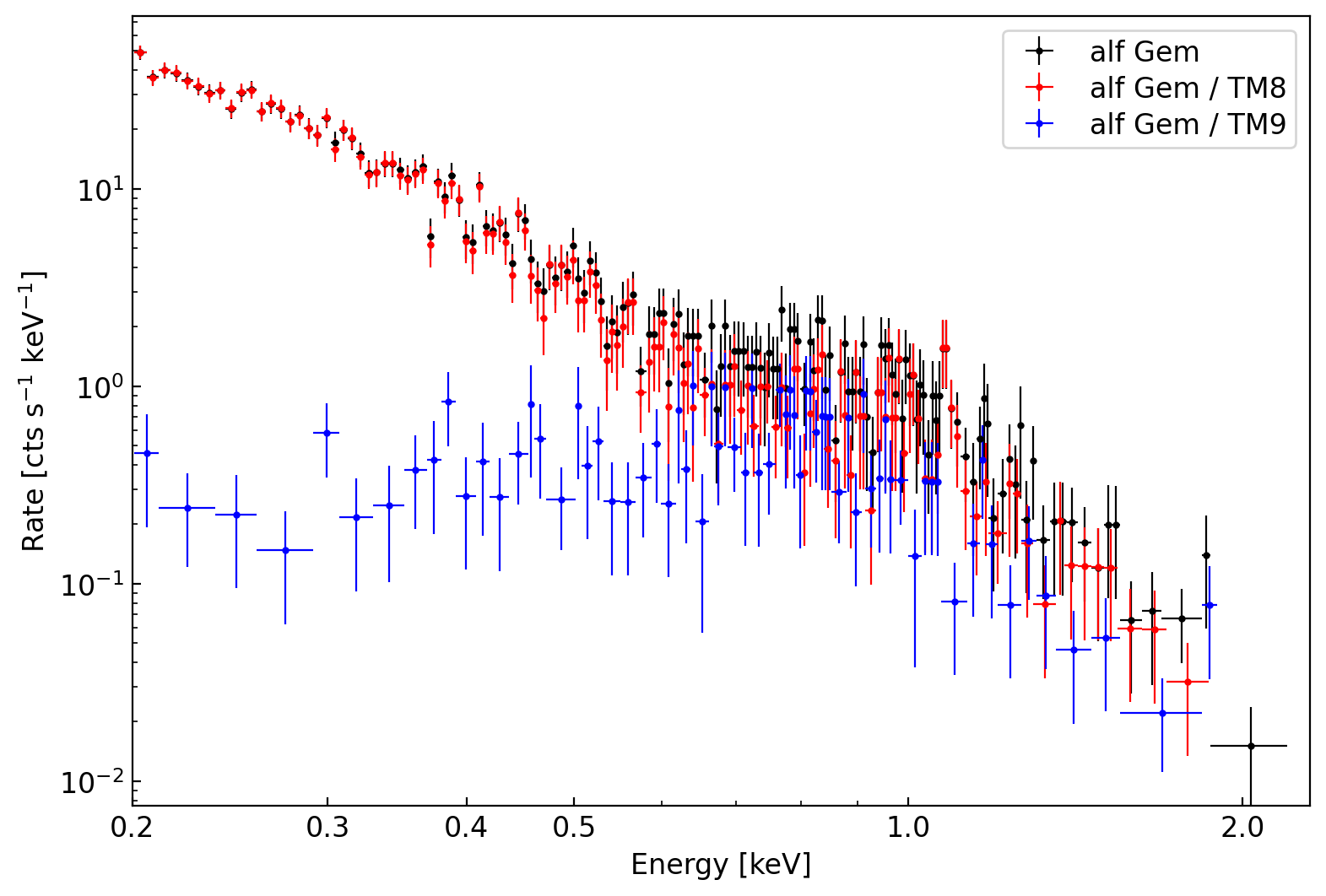}
\caption{\label{alfgem_tm}Spectrum and TM contributions for the optically bright X-ray source $\alpha$~Gem. The strong optical loading in TM8 is clearly visible.}
\end{figure}

\begin{figure}
\includegraphics[width=88mm]{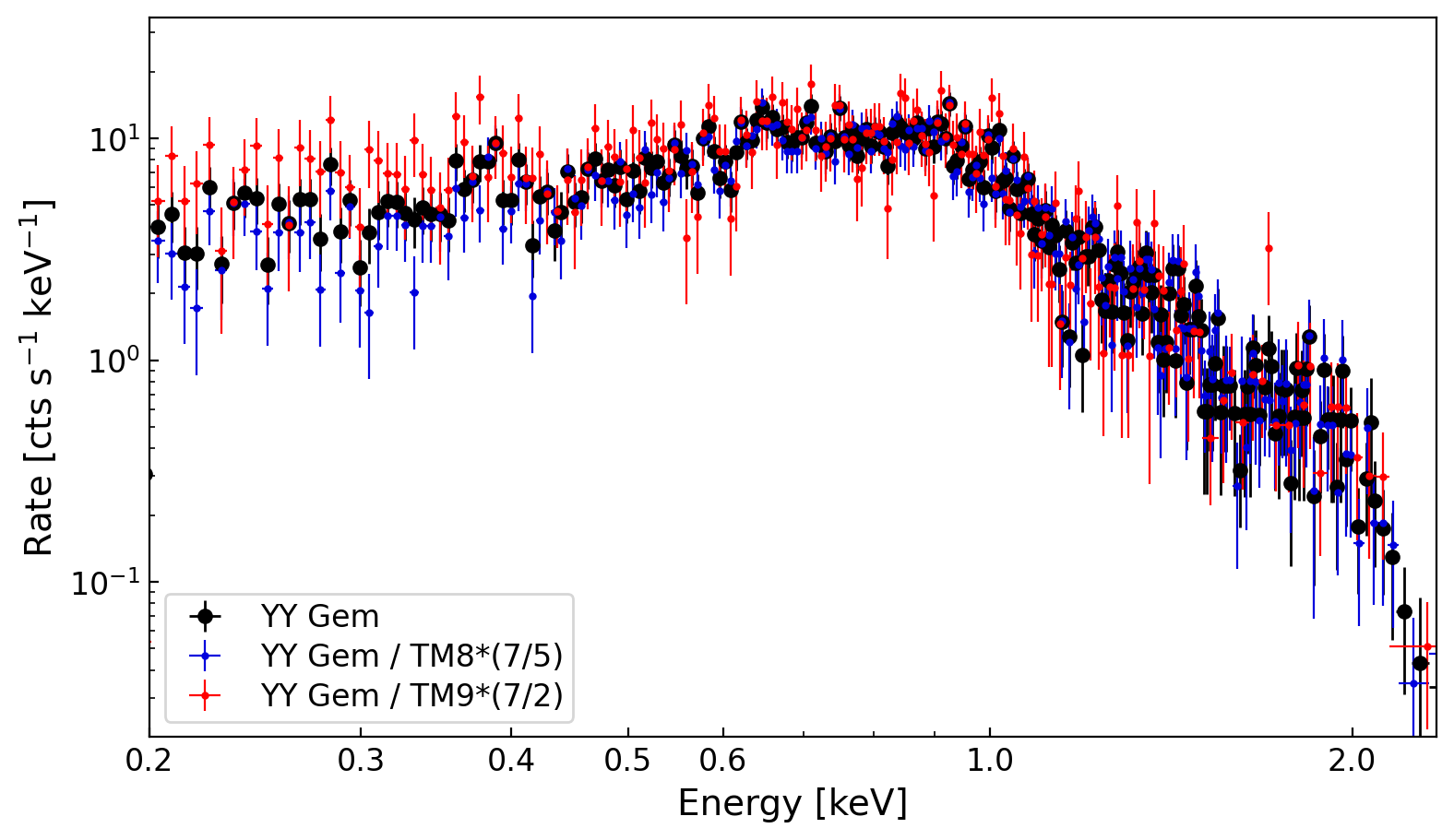}
\caption{\label{yy_tm2}Spectrum of the optically dark X-ray source YY~Gem with TM contributions.}
\end{figure}

\begin{figure}
\includegraphics[width=88mm]{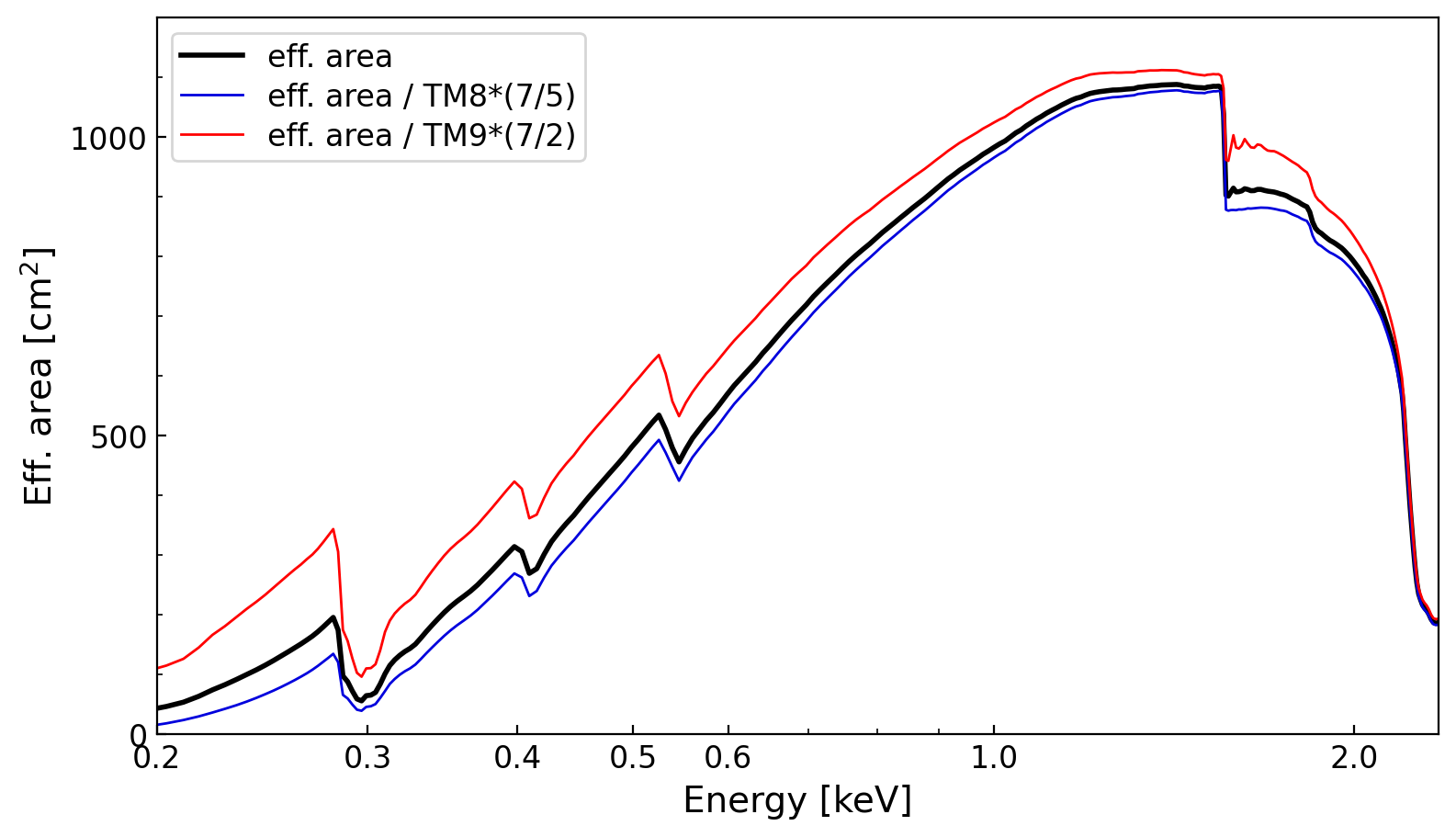}
\caption{\label{yy_eff}Survey effective area of eROSITA and TM contributions.}
\end{figure}

The so far discussed data refer to the full eROSITA instrument, but it is instructive to compare the spectra of telescope modules with on-chip filters (TM8) with those that have it in the filter wheel (TM9). Naively one would expect a greater sensitivity with TM9 due to their thinner Al layer, {yet also the poliyimide has substantial
blockage in the UV}.
 As shown in the top panel of Fig.~\ref{tmpat} for $\gamma$~TrA, the optical loading events are essentially exclusively detected in TM8. Very similar effects are seen, when the optically bright source is also an X-ray emitter. As shown for the case of $\alpha$~Gem in Fig.~\ref{alfgem_tm}, the optical loading contribution at softer energies in the TM8 modules is very pronounced, while basically absent in TM9. This behavior is different to real X-rays, where the spectra of TM8 and TM9 agree very well when considering their respective effective areas, as shown in Fig.\,\ref{yy_tm2} for the M dwarf binary YY~Gem. In Fig.\,\ref{yy_eff} we show the respective effective areas for the survey data.

When distinguishing the source events by their pattern, the relative contributions again mainly reflect the respective stellar brightness. {This can again
be recognised by inspecting the pattern distribution of the recorded events; we distinguish between 'single' events (denoted by 's' in Fig.~\ref{tmpat} bottom),
'double' events (denoted by 'd'), 'triple' events (denoted by 't') and 'quadruple' events (denoted by 'q').  An exemplary optical loading case
is shown for the star $\gamma$~TrA Fig.~\ref{tmpat} (bottom), where one sees that $\gamma$~TrA it is dominated by 'double' events, followed by 
'single' events and the higher patterns, i.e. summed 'triple' and 'quadruple' events, contribute by about 10\,\%, 
and are mainly found at energies above 0.3~keV. 
The respective pattern trend depends on optical brightness, in brighter stars the multiplicity of the events increases, in fainter ones the multiplicity 
decreases and 'single' events start to dominate at an optical brightness of around 3.5\,--\,4.0~mag.} The source events are largely found below 0.5~keV and higher energies and higher pattern fade out towards the fainter stars.

\section{eRASS - parameter and conversion factors}
\label{res_cal}

To categorize stellar X-ray sources, we determine hardness ratios and rate to flux conversions, i.e. 1/ECFs (energy conversion factor), for different plasma temperatures and absorption columns. Based on the measured count rates, they provide X-ray flux and basic spectral properties, when a spectral analysis is not suitable or desired. Parameter like hardness ratios also help to distinguish potentially optically generated or affected sources from those with predominant or exclusive X-ray emission.

\subsection{Coronae and thermal plasma}
\label{para_cor}

For stellar coronae we use single temperature plasma models on a grid, that covers the temperatures 0.1\,--\,10.0~keV (1~keV = 11.6~MK) and absorption values up to $N_{\rm H} = 1\times 10^{22}$\,cm$^{-2}$.
Here we focus on the main 0.2\,--\,2.3~keV eRASS band, that includes the spectral region most relevant for optical loading. An extended view that includes the P4 (2\,-\,5 keV) band, relative count rate contributions and rate conversions factors for non-negligible absorption is given in the appendix.

\begin{figure}
\includegraphics[width=88mm]{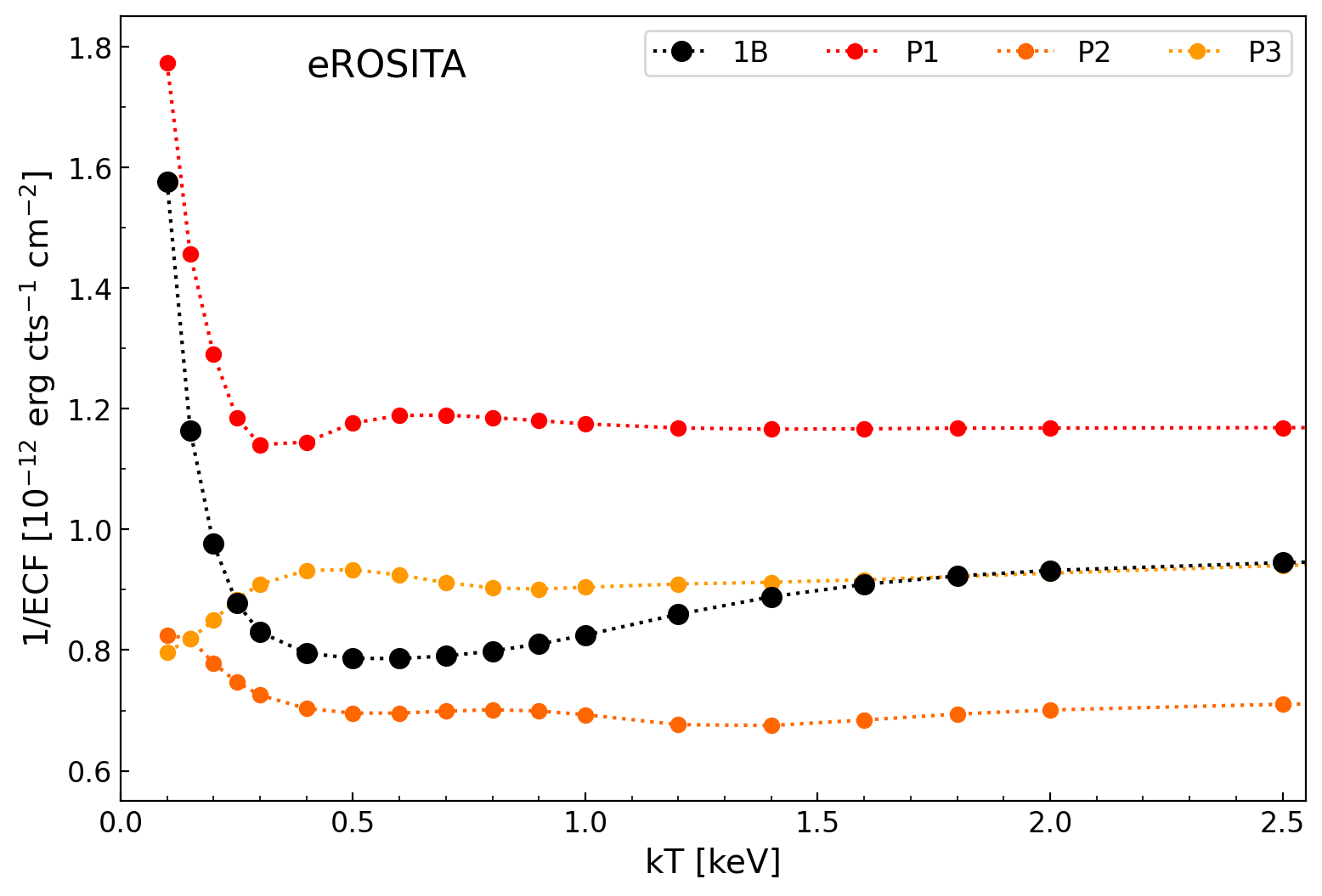}
\caption{\label{ecfs}Rate to flux conversion factors (1/ecf) for APEC models in the 0.2\,--\,2.5 keV energy range; legends 1B and P1-3 denote the shown eROSITA bands.}
\end{figure}

The calculated rate-to-flux conversion factors in the respective eROSITA bands are shown in Fig.~\ref{ecfs}. The given conversions apply to single temperature models, multi-temperature plasma can be approximated by its average temperature or by a combination of its contributions. One finds that values of $0.8-0.9 \times 10^{-12}$ erg\,cts$^{-1}$\,cm$^{-2}$ are an overall good choice for the rates in the 0.2\,--\,2.3~keV (1B) band, especially if one deals with active stars that make up the majority of stellar sources in the eRASS catalogs. The conversions are quite stable over a wide range of temperatures; stronger deviations are mainly present for plasma at very low temperatures. The photometric bands P1-3, i.e. those that encompass the 0.2\,--2.0\,keV energy range, show a very similar behavior and a conversion that is only moderately temperature dependent. As in the full energy band, an exception is the P1 band at temperatures of $\lesssim$~2\,MK, where a strong rise is present due to the declining effective area at lower energies.

\begin{figure}
\includegraphics[width=88mm]{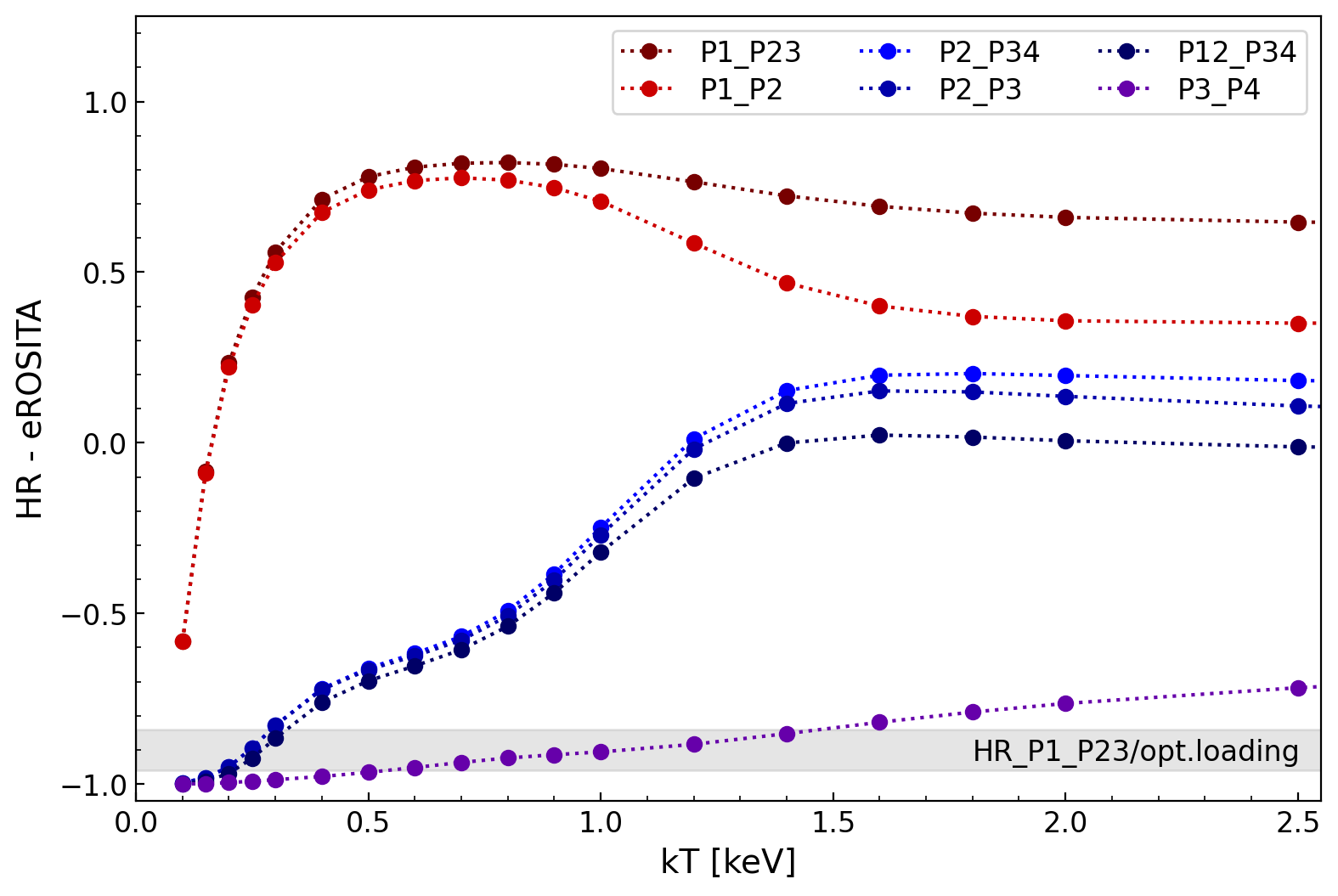}
\caption{\label{hrs}Hardness ratios defined as HR=(H-S)/(H+S) vs. plasma temperature. The S\_H legend denotes the used eROSITA bands. Values expected from optical loading are indicated as grey bar.}
\end{figure}

The hardness ratios HR\,=\,(H-S)/(H+S) in several eRASS energy bands are shown Fig.~\ref{hrs}, here e.g. P23 denotes the sum of count rates in the P2+P3 band. The P1\_P23 ratio is most suited to distinguish true X-rays from apparent X-ray events generated by optical loading. Values significantly below zero are typically not observed in at least moderately active stars with ${\rm log}~L_{\rm X}/L_{\rm bol} \gtrsim -5$ and with an average coronal temperatures above a few MK. Even inactive stars have a hardness ratio above -0.5, if their coronal temperature is around or above 1.5~MK, i.e. at temperatures of the Sun around activity minimum. As outlined below, optical loading generates hardness ratios well below values of typical coronal sources.

The observed X-ray emission may be subject to interstellar absorption and the ECFs have also been calculated for various $N_{\rm H}$ columns. In the 0.2\,--\,2.3~keV eROSITA energy band, X-ray absorption is quite minor and at a level of about 5\,\% for 1~keV plasma and $N_{\rm H} = 10^{20} {\rm cm}^{-2}$. It is therefore neglected in the analysis of nearby stars and the optical loading study.

\subsection{The case of optical loading}
\label{para_opt}

To estimate the hardness ratio from pure optical loading, we use the power-law models found in the spectral analysis as discussed in Sect.~\ref{res_spec}. For HR\_P1\_23 and similarly for HR\_P12, we get $HR = -0.90 \pm 0.03$ when using exponents in the range of alpha = -5.7 ... -6.5 and absorption of $N_{\rm H} \le 10^{20}$~cm$^{-2}$. These values are indicated with their two sigma range, shown as grey bar in Fig.~\ref{hrs}.

These very low hardness ratios are well below any expected coronal value and generally found only for stars with pure optical loading, as shown for the detected eRASS sources. If the source is instead optically bright and an intrinsic X-ray emitter, intermediate values are found in the combined models as well as in the eRASS data. In real data, noise and background contribute to the observed hardness ratios and errors become more significant, especially for the many fainter sources.

\section{Discussion and summary}
\label{sum}

We have studied optical loading in the eRASS data, specifically making use of the eRASS:3 catalog. Our main findings apply, potentially with appropriate scaling, to all eROSITA observations performed in survey or scanning mode. We show that optical loading and its effects are a complex interplay, but some universal criteria apply.

\begin{enumerate}

\item The optical loading flag 'FLAG\_OPT' is set for stars brighter than 5~mag (B, V, G) or 3.5 mag (J) for the eRASS:3 data release DR2 catalogs. This is a well motivated indicator, but it does not exclude optical loading for a non-flagged source, nor does it indicate significant impact if present.

\item Optical loading for bright and X-ray emitting stars can be roughly estimated, but accuracy suffers. Nevertheless, if the X-ray brightness to optical loading contrast is sufficiently large, basic X-ray properties can be obtained. Secondary parameters to consider are stellar color, potential variability and eRASS exposure depth and recipes for mitigation in moderately effected sources are provided. These include count rate corrections, dedicated spectral modeling or the analysis in an adapted spectral range.

\item Optical loading from very bright stars leads to saturation effects and severe distortions of the target properties. The eRASS data is hardly usable for most science cases and its calibration is undetermined. The pure optical loading sources show brightness dependent count rates and can be discerned by their very low hardness ratios.

\item Count rate conversions  are determined for thermal plasma emission in the 0.1\,--\,5.0~keV temperature range, that base on data simulated with APEC models. The derived values can be used for the analysis of stellar sources in general.

\end{enumerate}

\begin{acknowledgements}
This work is based on data from eROSITA, the soft X-ray instrument aboard SRG, a joint Russian-German science mission supported by the Russian Space Agency (Roskosmos), in the interests of the Russian Academy of Sciences represented by its Space Research Institute (IKI), and the Deutsches Zentrum f{\"u}r Luft- und Raumfahrt (DLR). The SRG spacecraft was built by Lavochkin Association (NPOL) and its subcontractors, and is operated by NPOL with support from the Max Planck Institute for Extraterrestrial Physics (MPE).

The development and construction of the eROSITA X-ray instrument was led by MPE, with contributions from the Dr. Karl Remeis Observatory Bamberg \& ECAP (FAU Erlangen-Nuernberg), the University of Hamburg Observatory, the Leibniz Institute for Astrophysics Potsdam (AIP), and the Institute for Astronomy and Astrophysics of the University of Tuebingen, with the support of DLR and the Max Planck Society. The Argelander Institute for Astronomy of the University of Bonn and the Ludwig Maximilians Universitaet Munich also participated in the science preparation for eROSITA. The eROSITA data shown here were processed using the eSASS software system developed by the German eROSITA consortium.\\

J.R. acknowledges support from the DLR under grant 50QR2505.
This research has made use of the SIMBAD database, operated at CDS, Strasbourg, France.
This work made use of the Astropy and PyXspec package.

\end{acknowledgements}

\bibliographystyle{aa}
\bibliography{erobib}

\begin{thebibliography}{16}
\expandafter\ifx\csname natexlab\endcsname\relax\def\natexlab#1{#1}\fi

\bibitem[{{Arnaud}(1996)}]{xspec}
{Arnaud}, K.~A. 1996, in ASP Conf.\ Ser, Vol. 101, Astronomical Data Analysis
  Software and Systems V, ed. G.~H. {Jacoby} \& J.~{Barnes} (SF: ASP), 17

\bibitem[{{Boller} {et~al.}(2016){Boller}, {Freyberg}, {Tr{\"u}mper}, {Haberl},
  {Voges}, \& {Nandra}}]{2RXS}
{Boller}, T., {Freyberg}, M.~J., {Tr{\"u}mper}, J., {et~al.} 2016, \aap, 588,
  A103

\bibitem[{{Brunner} {et~al.}(2022){Brunner}, {Liu}, {Lamer}, {Georgakakis},
  {Merloni}, {Brusa}, {Bulbul}, {Dennerl}, {Friedrich}, {Liu}, {Maitra},
  {Nandra}, {Ramos-Ceja}, {Sanders}, {Stewart}, {Boller}, {Buchner}, {Clerc},
  {Comparat}, {Dwelly}, {Eckert}, {Finoguenov}, {Freyberg}, {Ghirardini},
  {Gueguen}, {Haberl}, {Kreykenbohm}, {Krumpe}, {Osterhage}, {Pacaud},
  {Predehl}, {Reiprich}, {Robrade}, {Salvato}, {Santangelo}, {Schrabback},
  {Schwope}, \& {Wilms}}]{esass}
{Brunner}, H., {Liu}, T., {Lamer}, G., {et~al.} 2022, \aap, 661, A1

\bibitem[{{Cutri} {et~al.}(2003){Cutri}, {Skrutskie}, {van Dyk}, {Beichman},
  {Carpenter}, {Chester}, {Cambresy}, {Evans}, {Fowler}, {Gizis}, {Howard},
  {Huchra}, {Jarrett}, {Kopan}, {Kirkpatrick}, {Light}, {Marsh}, {McCallon},
  {Schneider}, {Stiening}, {Sykes}, {Weinberg}, {Wheaton}, {Wheelock}, \&
  {Zacarias}}]{2mass}
{Cutri}, R.~M., {Skrutskie}, M.~F., {van Dyk}, S., {et~al.} 2003, {2MASS All
  Sky Catalog of point sources.}

\bibitem[{{Gaia Collaboration} {et~al.}(2023){Gaia Collaboration}, {Vallenari},
  {Brown}, {Prusti}, {de Bruijne}, {Arenou}, {Babusiaux}, {Biermann},
  {Creevey}, {Ducourant}, {Evans}, {Eyer}, {Guerra}, {Hutton}, {Jordi},
  {Klioner}, {Lammers}, {Lindegren}, {Luri}, {Mignard}, {Panem}, {Pourbaix},
  {Randich}, {Sartoretti}, {Soubiran}, {Tanga}, {Walton}, {Bailer-Jones},
  {Bastian}, {Drimmel}, {Jansen}, {Katz}, {Lattanzi}, {van Leeuwen}, {Bakker},
  {Cacciari}, {Casta{\~n}eda}, {De Angeli}, {Fabricius}, {Fouesneau},
  {Fr{\'e}mat}, {Galluccio}, {Guerrier}, {Heiter}, {Masana}, {Messineo},
  {Mowlavi}, {Nicolas}, {Nienartowicz}, {Pailler}, {Panuzzo}, {Riclet}, {Roux},
  {Seabroke}, {Sordo}, {Th{\'e}venin}, {Gracia-Abril}, {Portell}, {Teyssier},
  {Altmann}, {Andrae}, {Audard}, {Bellas-Velidis}, {Benson}, {Berthier},
  {Blomme}, {Burgess}, {Busonero}, {Busso}, {C{\'a}novas}, {Carry}, {Cellino},
  {Cheek}, {Clementini}, {Damerdji}, {Davidson}, {de Teodoro}, {Nu{\~n}ez
  Campos}, {Delchambre}, {Dell'Oro}, {Esquej}, {Fern{\'a}ndez-Hern{\'a}ndez},
  {Fraile}, {Garabato}, {Garc{\'\i}a-Lario}, {Gosset}, {Haigron}, {Halbwachs},
  {Hambly}, {Harrison}, {Hern{\'a}ndez}, {Hestroffer}, {Hodgkin}, {Holl},
  {Jan{\ss}en}, {Jevardat de Fombelle}, {Jordan}, {Krone-Martins}, {Lanzafame},
  {L{\"o}ffler}, {Marchal}, {Marrese}, {Moitinho}, {Muinonen}, {Osborne},
  {Pancino}, {Pauwels}, {Recio-Blanco}, {Reyl{\'e}}, {Riello}, {Rimoldini},
  {Roegiers}, {Rybizki}, {Sarro}, {Siopis}, {Smith}, {Sozzetti}, {Utrilla},
  {van Leeuwen}, {Abbas}, {{\'A}brah{\'a}m}, {Abreu Aramburu}, {Aerts},
  {Aguado}, {Ajaj}, {Aldea-Montero}, {Altavilla}, {{\'A}lvarez}, {Alves},
  {Anders}, {Anderson}, {Anglada Varela}, {Antoja}, {Baines}, {Baker},
  {Balaguer-N{\'u}{\~n}ez}, {Balbinot}, {Balog}, {Barache}, {Barbato},
  {Barros}, {Barstow}, {Bartolom{\'e}}, {Bassilana}, {Bauchet}, {Becciani},
  {Bellazzini}, {Berihuete}, {Bernet}, {Bertone}, {Bianchi}, {Binnenfeld},
  {Blanco-Cuaresma}, {Blazere}, {Boch}, {Bombrun}, {Bossini}, {Bouquillon},
  {Bragaglia}, {Bramante}, {Breedt}, {Bressan}, {Brouillet}, {Brugaletta},
  {Bucciarelli}, {Burlacu}, {Butkevich}, {Buzzi}, {Caffau}, {Cancelliere},
  {Cantat-Gaudin}, {Carballo}, {Carlucci}, {Carnerero}, {Carrasco},
  {Casamiquela}, {Castellani}, {Castro-Ginard}, {Chaoul}, {Charlot}, {Chemin},
  {Chiaramida}, {Chiavassa}, {Chornay}, {Comoretto}, {Contursi}, {Cooper},
  {Cornez}, {Cowell}, {Crifo}, {Cropper}, {Crosta}, {Crowley}, {Dafonte},
  {Dapergolas}, {David}, {David}, {de Laverny}, {De Luise}, {De March}, {De
  Ridder}, {de Souza}, {de Torres}, {del Peloso}, {del Pozo}, {Delbo},
  {Delgado}, {Delisle}, {Demouchy}, {Dharmawardena}, {Di Matteo}, {Diakite},
  {Diener}, {Distefano}, {Dolding}, {Edvardsson}, {Enke}, {Fabre}, {Fabrizio},
  {Faigler}, {Fedorets}, {Fernique}, {Fienga}, {Figueras}, {Fournier},
  {Fouron}, {Fragkoudi}, {Gai}, {Garcia-Gutierrez}, {Garcia-Reinaldos},
  {Garc{\'\i}a-Torres}, {Garofalo}, {Gavel}, {Gavras}, {Gerlach}, {Geyer},
  {Giacobbe}, {Gilmore}, {Girona}, {Giuffrida}, {Gomel}, {Gomez},
  {Gonz{\'a}lez-N{\'u}{\~n}ez}, {Gonz{\'a}lez-Santamar{\'\i}a},
  {Gonz{\'a}lez-Vidal}, {Granvik}, {Guillout}, {Guiraud},
  {Guti{\'e}rrez-S{\'a}nchez}, {Guy}, {Hatzidimitriou}, {Hauser}, {Haywood},
  {Helmer}, {Helmi}, {Sarmiento}, {Hidalgo}, {Hilger}, {H{\l}adczuk}, {Hobbs},
  {Holland}, {Huckle}, {Jardine}, {Jasniewicz}, {Jean-Antoine Piccolo},
  {Jim{\'e}nez-Arranz}, {Jorissen}, {Juaristi Campillo}, {Julbe}, {Karbevska},
  {Kervella}, {Khanna}, {Kontizas}, {Kordopatis}, {Korn}, {K{\'o}sp{\'a}l},
  {Kostrzewa-Rutkowska}, {Kruszy{\'n}ska}, {Kun}, {Laizeau}, {Lambert},
  {Lanza}, {Lasne}, {Le Campion}, {Lebreton}, {Lebzelter}, {Leccia}, {Leclerc},
  {Lecoeur-Taibi}, {Liao}, {Licata}, {Lindstr{\o}m}, {Lister}, {Livanou},
  {Lobel}, {Lorca}, {Loup}, {Madrero Pardo}, {Magdaleno Romeo}, {Managau},
  {Mann}, {Manteiga}, {Marchant}, {Marconi}, {Marcos}, {Marcos Santos},
  {Mar{\'\i}n Pina}, {Marinoni}, {Marocco}, {Marshall}, {Martin Polo},
  {Mart{\'\i}n-Fleitas}, {Marton}, {Mary}, {Masip}, {Massari},
  {Mastrobuono-Battisti}, {Mazeh}, {McMillan}, {Messina}, {Michalik}, {Millar},
  {Mints}, {Molina}, {Molinaro}, {Moln{\'a}r}, {Monari}, {Mongui{\'o}},
  {Montegriffo}, {Montero}, {Mor}, {Mora}, {Morbidelli}, {Morel}, {Morris},
  {Muraveva}, {Murphy}, {Musella}, {Nagy}, {Noval}, {Oca{\~n}a}, {Ogden},
  {Ordenovic}, {Osinde}, {Pagani}, {Pagano}, {Palaversa}, {Palicio},
  {Pallas-Quintela}, {Panahi}, {Payne-Wardenaar}, {Pe{\~n}alosa Esteller},
  {Penttil{\"a}}, {Pichon}, {Piersimoni}, {Pineau}, {Plachy}, {Plum}, {Poggio},
  {Pr{\v{s}}a}, {Pulone}, {Racero}, {Ragaini}, {Rainer}, {Raiteri}, {Rambaux},
  {Ramos}, {Ramos-Lerate}, {Re Fiorentin}, {Regibo}, {Richards}, {Rios Diaz},
  {Ripepi}, {Riva}, {Rix}, {Rixon}, {Robichon}, {Robin}, {Robin}, {Roelens},
  {Rogues}, {Rohrbasser}, {Romero-G{\'o}mez}, {Rowell}, {Royer}, {Ruz Mieres},
  {Rybicki}, {Sadowski}, {S{\'a}ez N{\'u}{\~n}ez}, {Sagrist{\`a} Sell{\'e}s},
  {Sahlmann}, {Salguero}, {Samaras}, {Sanchez Gimenez}, {Sanna},
  {Santove{\~n}a}, {Sarasso}, {Schultheis}, {Sciacca}, {Segol}, {Segovia},
  {S{\'e}gransan}, {Semeux}, {Shahaf}, {Siddiqui}, {Siebert}, {Siltala},
  {Silvelo}, {Slezak}, {Slezak}, {Smart}, {Snaith}, {Solano}, {Solitro},
  {Souami}, {Souchay}, {Spagna}, {Spina}, {Spoto}, {Steele},
  {Steidelm{\"u}ller}, {Stephenson}, {S{\"u}veges}, {Surdej}, {Szabados},
  {Szegedi-Elek}, {Taris}, {Taylor}, {Teixeira}, {Tolomei}, {Tonello}, {Torra},
  {Torra}, {Torralba Elipe}, {Trabucchi}, {Tsounis}, {Turon}, {Ulla}, {Unger},
  {Vaillant}, {van Dillen}, {van Reeven}, {Vanel}, {Vecchiato}, {Viala},
  {Vicente}, {Voutsinas}, {Weiler}, {Wevers}, {Wyrzykowski}, {Yoldas}, {Yvard},
  {Zhao}, {Zorec}, {Zucker}, \& {Zwitter}}]{gdr3}
{Gaia Collaboration}, {Vallenari}, A., {Brown}, A.~G.~A., {et~al.} 2023, \aap,
  674, A1

\bibitem[{{Grevesse} \& {Sauval}(1998)}]{grsa}
{Grevesse}, N. \& {Sauval}, A.~J. 1998, \ssr, 85, 161

\bibitem[{{Hoffleit} \& {Jaschek}(1991)}]{bsc91}
{Hoffleit}, D. \& {Jaschek}, C. 1991, {The Bright star catalogue}

\bibitem[{{H{\o}g} {et~al.}(2000){H{\o}g}, {Fabricius}, {Makarov}, {Urban},
  {Corbin}, {Wycoff}, {Bastian}, {Schwekendiek}, \& {Wicenec}}]{tycho2}
{H{\o}g}, E., {Fabricius}, C., {Makarov}, V.~V., {et~al.} 2000, \aap, 355, L27

\bibitem[{{Merloni} {et~al.}(2024){Merloni}, {Lamer}, {Liu}, {Ramos-Ceja},
  {Brunner}, {Bulbul}, {Dennerl}, {Doroshenko}, {Freyberg}, {Friedrich},
  {Gatuzz}, {Georgakakis}, {Haberl}, {Igo}, {Kreykenbohm}, {Liu}, {Maitra},
  {Malyali}, {Mayer}, {Nandra}, {Predehl}, {Robrade}, {Salvato}, {Sanders},
  {Stewart}, {Tub{\'\i}n-Arenas}, {Weber}, {Wilms}, {Arcodia}, {Artis},
  {Aschersleben}, {Avakyan}, {Aydar}, {Bahar}, {Balzer}, {Becker}, {Berger},
  {Boller}, {Bornemann}, {Br{\"u}ggen}, {Brusa}, {Buchner}, {Burwitz},
  {Camilloni}, {Clerc}, {Comparat}, {Coutinho}, {Czesla}, {Dannhauer},
  {Dauner}, {Dauser}, {Dietl}, {Dolag}, {Dwelly}, {Egg}, {Ehl}, {Freund},
  {Friedrich}, {Gaida}, {Garrel}, {Ghirardini}, {Gokus}, {Gr{\"u}nwald},
  {Grandis}, {Grotova}, {Gruen}, {Gueguen}, {H{\"a}mmerich}, {Hamaus},
  {Hasinger}, {Haubner}, {Homan}, {Ider Chitham}, {Joseph}, {Joyce},
  {K{\"o}nig}, {Kaltenbrunner}, {Khokhriakova}, {Kink}, {Kirsch}, {Kluge},
  {Knies}, {Krippendorf}, {Krumpe}, {Kurpas}, {Li}, {Liu}, {Locatelli},
  {Lorenz}, {M{\"u}ller}, {Magaudda}, {Mannes}, {McCall}, {Meidinger},
  {Michailidis}, {Migkas}, {Mu{\~n}oz-Giraldo}, {Musiimenta}, {Nguyen-Dang},
  {Ni}, {Olechowska}, {Ota}, {Pacaud}, {Pasini}, {Perinati}, {Pires},
  {Pommranz}, {Ponti}, {Poppenhaeger}, {P{\"u}hlhofer}, {Rau}, {Reh},
  {Reiprich}, {Roster}, {Saeedi}, {Santangelo}, {Sasaki}, {Schmitt},
  {Schneider}, {Schrabback}, {Schuster}, {Schwope}, {Seppi}, {Serim},
  {Shreeram}, {Sokolova-Lapa}, {Starck}, {Stelzer}, {Stierhof}, {Suleimanov},
  {Tenzer}, {Traulsen}, {Tr{\"u}mper}, {Tsuge}, {Urrutia}, {Veronica},
  {Waddell}, {Willer}, {Wolf}, {Yeung}, {Zainab}, {Zangrandi}, {Zhang},
  {Zhang}, \& {Zheng}}]{DR1}
{Merloni}, A., {Lamer}, G., {Liu}, T., {et~al.} 2024, \aap, 682, A34

\bibitem[{{Predehl} {et~al.}(2021){Predehl}, {Andritschke}, {Arefiev},
  {Babyshkin}, {Batanov}, {Becker}, {B{\"o}hringer}, {Bogomolov}, {Boller},
  {Borm}, {Bornemann}, {Br{\"a}uninger}, {Br{\"u}ggen}, {Brunner}, {Brusa},
  {Bulbul}, {Buntov}, {Burwitz}, {Burkert}, {Clerc}, {Churazov}, {Coutinho},
  {Dauser}, {Dennerl}, {Doroshenko}, {Eder}, {Emberger}, {Eraerds},
  {Finoguenov}, {Freyberg}, {Friedrich}, {Friedrich}, {F{\"u}rmetz},
  {Georgakakis}, {Gilfanov}, {Granato}, {Grossberger}, {Gueguen}, {Gureev},
  {Haberl}, {H{\"a}lker}, {Hartner}, {Hasinger}, {Huber}, {Ji}, {Kienlin},
  {Kink}, {Korotkov}, {Kreykenbohm}, {Lamer}, {Lomakin}, {Lapshov}, {Liu},
  {Maitra}, {Meidinger}, {Menz}, {Merloni}, {Mernik}, {Mican}, {Mohr},
  {M{\"u}ller}, {Nandra}, {Nazarov}, {Pacaud}, {Pavlinsky}, {Perinati},
  {Pfeffermann}, {Pietschner}, {Ramos-Ceja}, {Rau}, {Reiffers}, {Reiprich},
  {Robrade}, {Salvato}, {Sanders}, {Santangelo}, {Sasaki}, {Scheuerle},
  {Schmid}, {Schmitt}, {Schwope}, {Shirshakov}, {Steinmetz}, {Stewart},
  {Str{\"u}der}, {Sunyaev}, {Tenzer}, {Tiedemann}, {Tr{\"u}mper}, {Voron},
  {Weber}, {Wilms}, \& {Yaroshenko}}]{erosita}
{Predehl}, P., {Andritschke}, R., {Arefiev}, V., {et~al.} 2021, \aap, 647, A1

\bibitem[{{Ramos-Ceja} {et~al.}(2026){Ramos-Ceja}, {Lamer}, {Salvato}, \&
  {more}}]{DR2}
{Ramos-Ceja}, M., {Lamer}, G., {Salvato}, M., \& {more}, X. 2026, \aap

\bibitem[{{Sasaki} {et~al.}(2024){Sasaki}, {Robrade}, {Krause}, {Knies},
  {Tsuge}, {P{\"u}hlhofer}, \& {Strong}}]{2024A&A...682A.172S}
{Sasaki}, M., {Robrade}, J., {Krause}, M. G.~H., {et~al.} 2024, \aap, 682, A172

\bibitem[{{Smith} {et~al.}(2001){Smith}, {Brickhouse}, {Liedahl}, \&
  {Raymond}}]{apec}
{Smith}, R.~K., {Brickhouse}, N.~S., {Liedahl}, D.~A., \& {Raymond}, J.~C.
  2001, \apjl, 556, L91

\bibitem[{{Stelzer} \& {Burwitz}(2003)}]{stelzer03}
{Stelzer}, B. \& {Burwitz}, V. 2003, \aap, 402, 719

\bibitem[{{Sunyaev} {et~al.}(2021){Sunyaev}, {Arefiev}, {Babyshkin},
  {Bogomolov}, {Borisov}, {Buntov}, {Brunner}, {Burenin}, {Churazov},
  {Coutinho}, {Eder}, {Eismont}, {Freyberg}, {Gilfanov}, {Gureyev}, {Hasinger},
  {Khabibullin}, {Kolmykov}, {Komovkin}, {Krivonos}, {Lapshov}, {Levin},
  {Lomakin}, {Lutovinov}, {Medvedev}, {Merloni}, {Mernik}, {Mikhailov},
  {Molodtsov}, {Mzhelsky}, {M{\"u}ller}, {Nandra}, {Nazarov}, {Pavlinsky},
  {Poghodin}, {Predehl}, {Robrade}, {Sazonov}, {Scheuerle}, {Shirshakov},
  {Tkachenko}, \& {Voron}}]{srg}
{Sunyaev}, R., {Arefiev}, V., {Babyshkin}, V., {et~al.} 2021, \aap, 656, A132

\bibitem[{{Voges} {et~al.}(1999){Voges}, {Aschenbach}, {Boller},
  {Br{\"a}uninger}, {Briel}, {Burkert}, {Dennerl}, {Englhauser}, {Gruber},
  {Haberl}, {Hartner}, {Hasinger}, {K{\"u}rster}, {Pfeffermann}, {Pietsch},
  {Predehl}, {Rosso}, {Schmitt}, {Tr{\"u}mper}, \& {Zimmermann}}]{1RXS}
{Voges}, W., {Aschenbach}, B., {Boller}, T., {et~al.} 1999, \aap, 349, 389

\end{thebibliography}

\begin{appendix}

\section{}
\label{app}
\begin{figure}
\includegraphics[width=88mm]{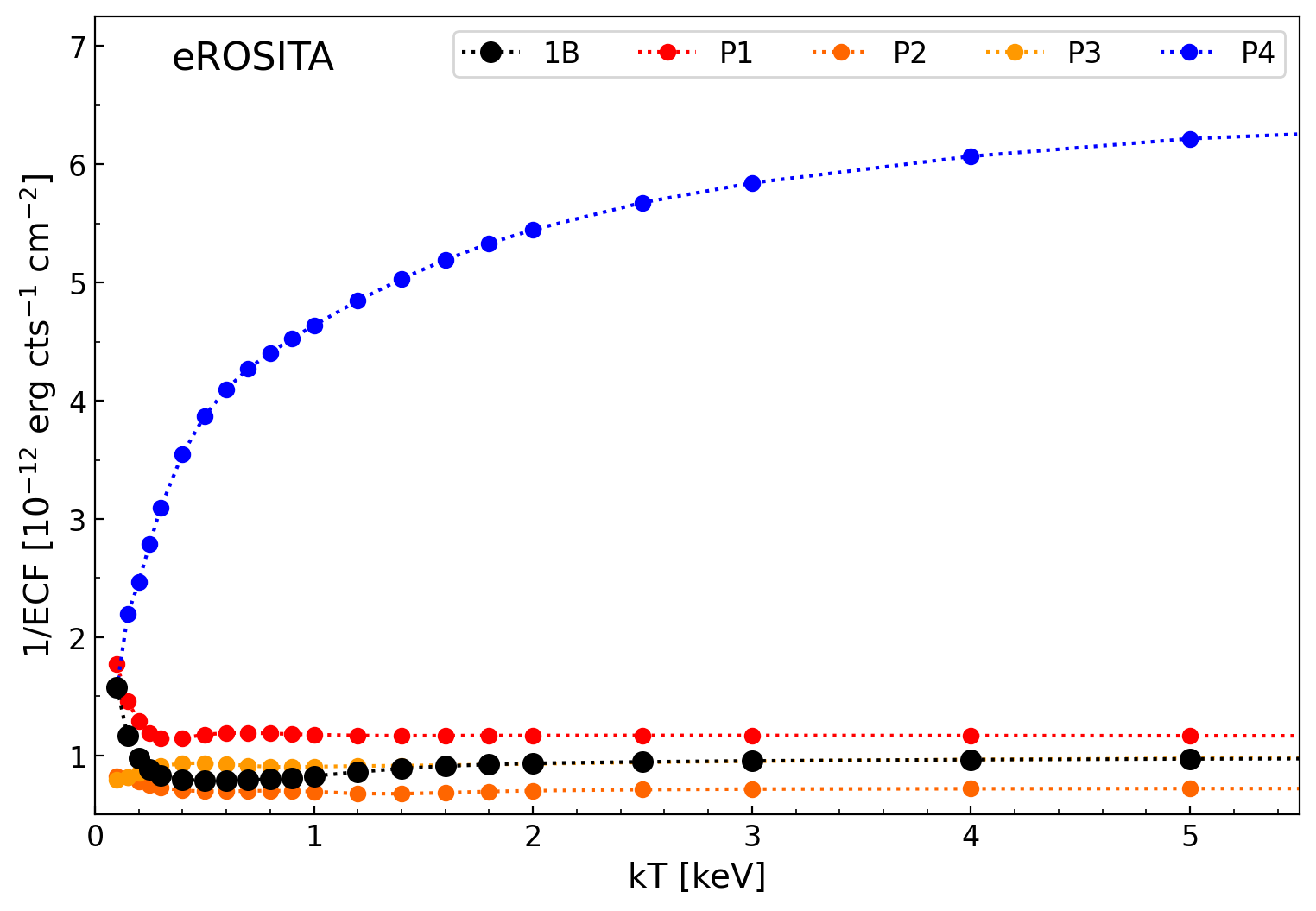}
\caption{\label{a_ecfs}Rate to flux conversion factors for APEC models; legend denotes the respective eROSITA energy bands.}
\end{figure}

\begin{figure}
\includegraphics[width=88mm]{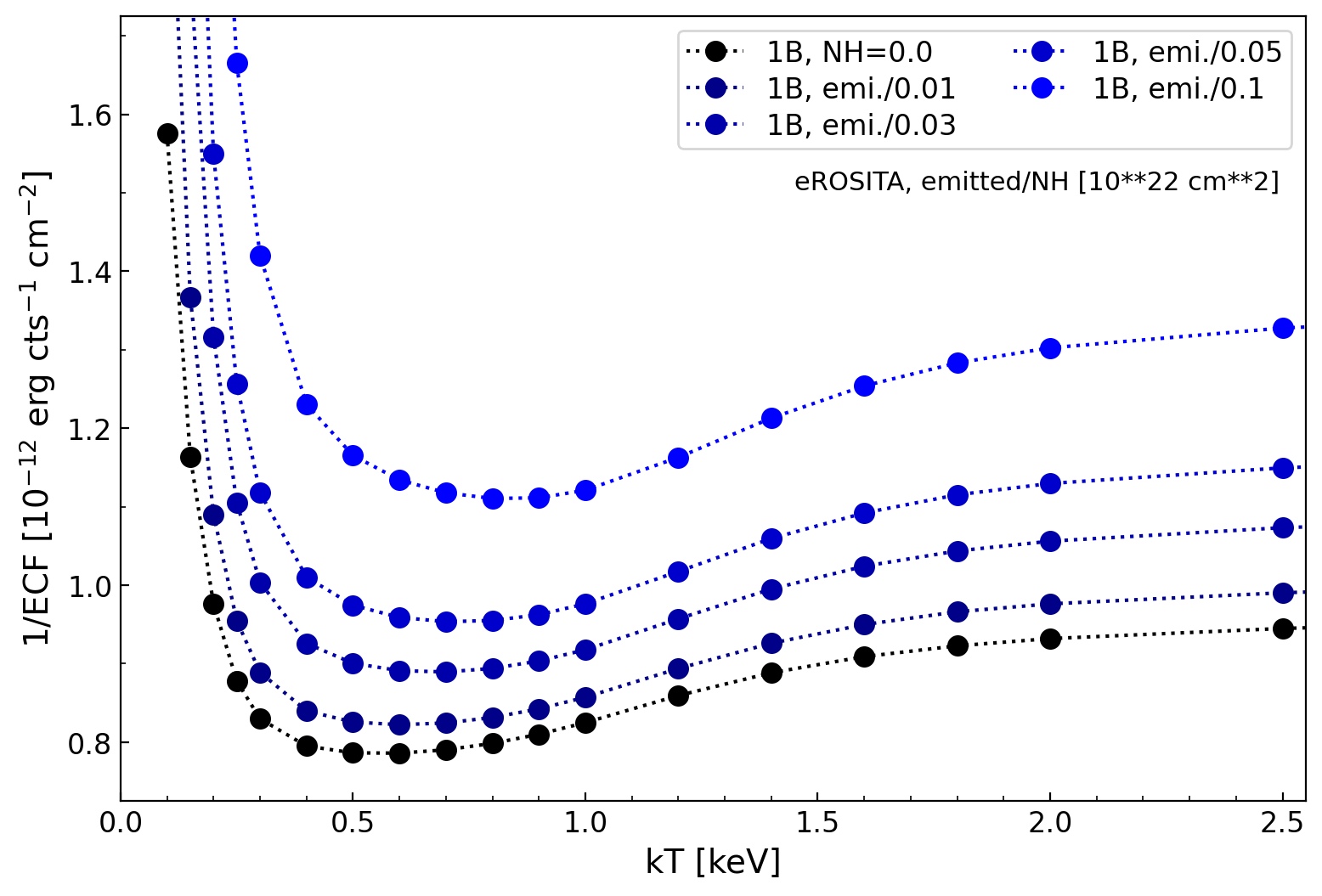}
\caption{\label{a_ecfs_nh}Rate to emitted flux conversion factors vs. plasma temperature for different column densities; 0.2\,--\,2.3~keV energy range; legend denotes column strength.}
\end{figure}

\begin{figure}
\includegraphics[width=88mm]{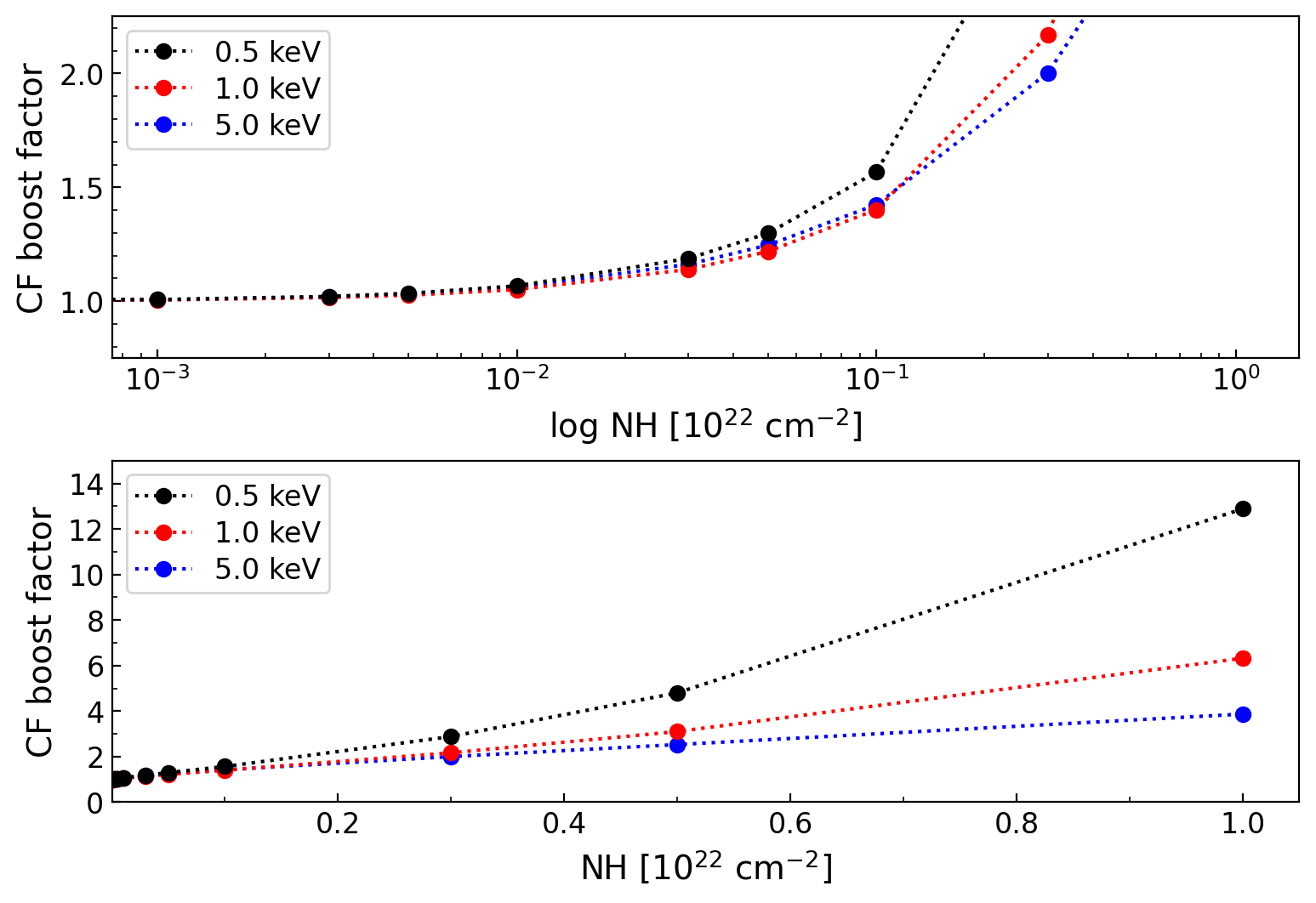}
\caption{\label{a_ecfs_nh1}Emitted flux; boost of rate conversion factor vs. absorption. {\it Top:} The factor is very moderate up to log\,NH$\approx$20 and around 1.5 at log\,NH=21. {\it Bottom:} Above log\,NH$\approx$21 the factor rises strongly and becomes more plasma temperature dependent.}
\end{figure}

\begin{figure}
\includegraphics[width=88mm]{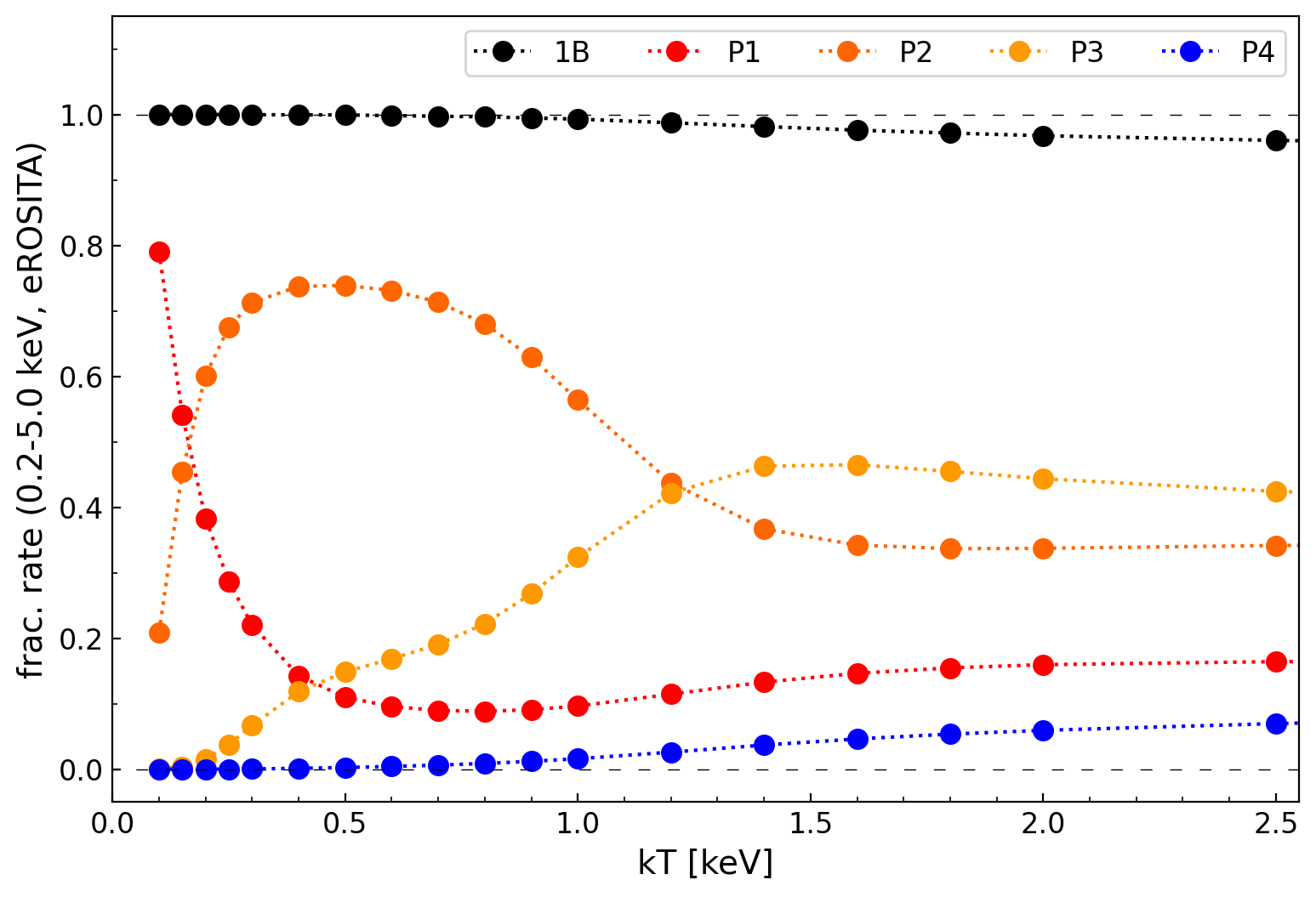}
\caption{\label{a_frc}Fractional rate contributions to the total 0.2\,--\,5.0 keV count rate; legend denotes the eROSITA energy bands.}
\end{figure}

eRASS performance parameters in addition to Sect.~\ref{res_cal}. Shown are conversion factors in the 0.2\,--\,5.0~keV energy range (\ref{a_ecfs}), the impact and correction taking into account absorption (\ref{a_ecfs_nh}, \ref{a_ecfs_nh1}) and the relative rate contributions from the eROSITA energy bands (\ref{a_frc}), namely 0.2\,--\,2.3~keV (1B) and 0.2\,--\,0.5\,--\,1.0\,--\,2.0\,--\,5.0~keV (P1-4).

\end{appendix}

\end{document}